\documentclass[longauth]{aa}  

\usepackage{graphicx}
\usepackage{txfonts}
\usepackage{lipsum}
\usepackage{subcaption}        
\usepackage{caption}
\usepackage{lscape}             
\usepackage{placeins}           
\usepackage{colortbl}
\usepackage[table]{xcolor}
\newcommand{\kms}{\ensuremath{\mathrm{km\ s}^{-1}}}

\usepackage[bookmarks=false]{hyperref}

\begin{document}

   \title{Complex Organic Molecules in Protostars with ALMA Spectral Surveys (COMPASS)}

   \subtitle{II. Approach to data reduction and products for line-rich broadband (sub)millimeter spectra}


   \author{Adele L. Plunkett \inst{1}\fnmsep\thanks{Corresponding author: aplunket@nrao.edu}
        \and S{\'e}bastien Maret\inst{2}
        \and Daniel Harsono\inst{3}
        \and Jes K. J{\o}rgensen\inst{4}
        \and Audrey Coutens\inst{5}
        \and Maria N. Drozdovskaya\inst{6,7}
        \and Jeong-Eun Lee\inst{8}
        \and Arnaud Belloche\inst{9}
        \and Fernando Cruz-S\'aenz de Miera \inst{5}
        \and Merel L.~R. van 't Hoff \inst{10}
        \and Jae-Hong Jeong \inst{8}
        \and Chul-Hwan Kim\inst{8}
        \and \'Agnes K\'osp\'al \inst{11,12,13}
        \and Martine L{\"u}tzen \inst{4}
        \and Brett A. McGuire\inst{14,1}
        \and Pooneh Nazari\inst{15}
        \and Silvia Spezzano\inst{16}
        \and Zoie Telkamp\inst{17}
        \and Yao-Lun Yang\inst{18}
        \and Hyeong-Sik Yun\inst{8,22}
        \and Audrey Andreu\inst{5}
        \and Jennifer Bergner\inst{19}
        \and Judit Ferrer Asensio\inst{18}
        \and Niels F.~W. Ligterink\inst{20}
        \and Yuxin Lin\inst{16}
        \and Sheng-Yuan Liu\inst{21}
        \and Pierre Marchand\inst{5}
        \and Mihika Rao\inst{17}
        \and Shaoshan Zeng\inst{18}
        }

   \institute{National Radio Astronomy Observatory, 520 Edgemont Road, Charlottesville, VA 22903, USA
   \and Univ. Grenoble Alpes, CNRS, IPAG, 38000 Grenoble, France
   \and Institute of Astronomy, Department of Physics, National Tsinghua University, No. 101, Section 2, Kuang-Fu Road, Hsinchu 30013, Taiwan, R.O.C
   \and Niels Bohr Institute, University of Copenhagen, Jagtvej 155A, DK-2200 Copenhagen N, Denmark
   \and Institut de Recherche en Astrophysique et Plan\'etologie, Universit\'e de Toulouse, CNRS, CNES, 9 av. du Colonel Roche, 31028 Toulouse Cedex 4, France
   \and Physikalisch-Meteorologisches Observatorium Davos und Weltstrahlungszentrum (PMOD/WRC), Dorfstrasse 33, 7260 Davos Dorf, Switzerland
   \and Department of Chemistry, Biochemistry and Pharmaceutical Sciences (DCBP), Universit{\"a}t Bern, Freiestrasse 3, 3012 Bern, Switzerland
   \and Department of Physics and Astronomy, Seoul National University,	1 Gwanak-ro, Gwanak-gu, Seoul 08826, Republic of Korea
    \and Max-Planck-Institut f{\"u}r Radioastronomie,	Auf dem H{\"u}gel 69, 53121 Bonn, Germany
    \and Purdue University, 525 Northwestern Ave, West Lafayette, IN 47907 USA
    \and Konkoly Observatory, HUN-REN Research Centre for Astronomy and Earth Sciences, MTA Centre of Excellence, Konkoly-Thege Mikl\'os \'ut 15-17, 1121 Budapest, Hungary
    \and Institute of Physics and Astronomy, ELTE E\"otv\"os Lor\'and University, P\'azm\'any P\'eter s\'et\'any 1/A, 1117 Budapest, Hungary
    \and Max-Planck-Insitut f\"ur Astronomie, K\"onigstuhl 17, 69117 Heidelberg, Germany
    \and Department of Chemistry, Massachusetts Institute of Technology, 77 Massachusetts Ave, Cambridge, MA 02139, USA
    \and European Southern Observatory, Karl-Schwarzschild-Strasse 2, 85748 Garching, Germany
    \and Max-Planck-Institut f{\"u}r extraterrestrische Physik, Giessenbachstrasse 1, 85748 Garching bei M{\"u}nchen, Germany
    \and Department of Astronomy, University of Virginia, 530 McCormick Road, Charlottesville, VA 22904, USA
    \and Star and Planet Formation Laboratory, RIKEN Pioneering Research Institute, Wako-shi, Saitama, 351-0198, Japan
    \and UC Berkeley, Department of Chemistry, Berkeley, CA 94720, USA
    \and Faculty of Aerospace Engineering, Delft University of Technology, Kluyverweg 1, 2629 HS Delft, The Netherlands
    \and Institute of Astronomy and Astrophysics, Academia Sinica,	11F of AS/NTU Astronomy-Mathematics Building, No.1, Sec. 4, Roosevelt Rd, Taipei 10617, Taiwan, R.O.C.
    \and Korea Astronomy and Space Science Institute, 776 Daedeok-daero, Yuseong-gu, Daejeon 34055, Republic of Korea
   }

\date{Accepted 20/04/2026}
 
  \abstract
   {Studying the chemical composition of the warm gas surrounding embedded solar-type protostars is one of the areas for which the Atacama Large Millimeter/submillimeter Array (ALMA) is particularly well suited.  Recently, the ALMA Large
Program \textit{Complex Organic Molecules in Protostars with ALMA Spectral Surveys (COMPASS)} carried out an observing campaign of a sample of 11 protostellar regions to investigate the chemical impact of their environments and evolutionary
stages.}
   {The aim of this paper is to provide the observational and procedural information necessary for the community to understand the analysis presented in forthcoming scientific papers by the COMPASS collaboration, and to enable archival research using the data products provided by our team and the ALMA Science Archive.} 
   {This paper describes the technical setup of the observations, and it outlines the data-processing pipeline and primary data products from this initiative.  The Common Astronomy Software Applications (CASA) was used for initial calibration following the standard ALMA Pipeline, and then IMAGER was used for self-calibration and imaging.  We describe post-processing done with Python to achieve favorable continuum subtraction, correct for the primary beam response, and extract spectra.}
   {We discuss our strategy for mitigating data challenges, including those related to data volume and continuum subtraction for spectral line analysis.  The incorporation of IMAGER in our workflow was advantageous in terms of the computing efficiency which allowed us to benchmark several imaging parameters and techniques; we also demonstrate that the imaging outcomes using IMAGER or CASA are scientifically comparable. Numerous data products are generated for the program, including image cubes for the 36 spectral windows targeting each source, extracted spectra, and continuum images. }
   {}

\titlerunning{COMPASS Data Reduction}
\authorrunning{Plunkett, A.~L. et al.}
\maketitle
\nolinenumbers

\section{Introduction} \label{sec:intro}
Understanding the origin and evolution of the complex organic molecules (COMs) observed in the interstellar medium is a fundamental goal in astrochemistry \citep[e.g.,][]{herbst2009,jorgensen2020}. 
During star and planet formation, the density, temperature, and ultra-violet (UV) flux can change by orders of magnitude, prompting the chemical reactions that yield a wide variety of molecules and their corresponding molecular emission. The chemical inventory of young stellar objects also reflects the physical properties of the surrounding protostellar environment. 

Hundreds of molecules have been detected to date in interstellar and circumstellar environments \citep{mcguire2022}, and a valuable strategy for disentangling chemical complexity is the execution of unbiased, large spectral line surveys. Such surveys allow for the systematic identification of molecular species and the determination of their column densities and rotational temperatures through numerous rotational transitions in the (sub)millimeter regime. While early single-dish surveys established the presence of complex chemistry in high-mass and/or bright sources \citep[e.g.,][]{johansson1984,blake1987,schilke1997,vandishoeckblake1998}, interferometric observations are crucial to distinguish the physical components driving this emission and to connect the chemical inventory to the evolutionary processes and physical conditions of low-mass protostars.

In this context, the Atacama Large Millimeter/submillimeter Array (ALMA) in Chile has clearly revolutionized the field with unbiased surveys of a range of different objects. 
Notable benchmarks include the Protostellar Interferometric Line Survey \citep[PILS;][]{jorgensen2016} targeting the low-mass hot corino IRAS~16293-2422 (329.1--362.9~GHz), the ALMA Spectral Survey of an Eruptive Young Star (ASSAY; \citealt{lee2024,jeong2025}) of the disk around V883~Ori (220.7--274.9~GHz), the G31.41+0.31 Unbiased ALMA sPectral Observational Survey (GUAPOS, 86.1--115.9 GHz; \citealt{mininni2020,lopezgallifa2024}), and the Exploring Molecular Complexity with ALMA (EMoCA) and Re-exploring Molecular Complexity with ALMA (ReMoCA) programs \citep{Belloche2016,Belloche2019} surveying the high-mass protocluster Sgr B2(N) in the Galactic center region (84.1--114.4 GHz).  These programs demonstrate the immense diagnostic power of high-sensitivity, wide-bandwidth interferometry. 

However, they also highlight the significant observational and computational challenges inherent to such datasets. 
In the case of COMs (carbon-bearing molecules with $\geq$6 atoms), the extremely line-rich spectra often reach the spectral confusion limit \cite[e.g.,][]{herbst2009}.  
In this regime, identifying weak molecular signals requires not only high sensitivity, but also robust strategies to disentangle contributions from thousands of transitions and accurately subtract the dust continuum emission.

Scaling these detailed studies from single sources to larger samples presents a major hurdle. Obtaining data of sufficient quality for astrochemical studies of protostars (sub-arcsecond angular resolution, a velocity resolution of a few tenths of km s$^{-1}$, and a sensitivity of a few mJy beam$^{-1}$) for a statistical sample requires substantial observing time (on the order of $\sim100$ hours) and results in massive data volumes ($\sim1$\ TB of raw data and $\sim100$\ GB of image products per source).
Processing such data is computationally intensive; high spectral resolution over broad bandwidths generates distinct challenges for continuum subtraction, imaging, and quality assurance, particularly when standard pipelines may struggle with line-confusion-limited fields.

This paper presents the technical implementation and data reduction approach for the ALMA Large Program (LP) Complex Organic Molecules in Protostars with ALMA Spectral Surveys (COMPASS), which was initiated during ALMA's Cycle 9 and completed during Cycle 10. 
COMPASS targets eleven protostellar regions in the frequency range of 279 to 312 GHz, with angular ($\lesssim0.5^{\prime\prime}$) and spectral ($\lesssim0.5~\textrm{km s}^{-1}$) resolutions designed to target science cases comparable and/or complementary to that of PILS.
A full description of the scientific motivation, observational design, and source sample of COMPASS is provided by \citet[][hereafter Paper I]{jorgensen2026}; readers can also refer to \citet{coutens2026} for an application of the spectral analysis from this program. The purpose of the present work is to document data-specific aspects of the program, from the raw telescope data, through calibration and imaging, to the delivery of enhanced data products.  We aim to provide a reference for the reduction techniques relevant to the first scientific results presented in Paper I and future COMPASS publications, as well as a reference for the wider community to utilize the LP's datasets. 
Furthermore, the techniques discussed here address common challenges in modern astrochemistry, and they are applicable to other line-rich interferometric studies, especially in the upcoming era of the ALMA Wideband Sensitivity Upgrade (WSU).

The structure of the paper is the following.  
In Sect.~\ref{sec:observations}, we describe the observations and the standard data products delivered by the ALMA Pipeline. 
In Sect.~\ref{sec:dataproc}, we describe the specialized calibration and imaging workflows developed by our team, the generation of enhanced data products, and the quality control measures.  
We close with a summary in Sect.~\ref{sec:conc}.  
An Appendix is provided for more specific information on the observations and an additional imaging procedure for reference.

\section{Observations} \label{sec:observations}
\begin{figure}[hbt]
\begin{minipage}{0.48\textwidth} 
\includegraphics[width=\textwidth, trim=2cm 0cm 2cm 2cm, clip]{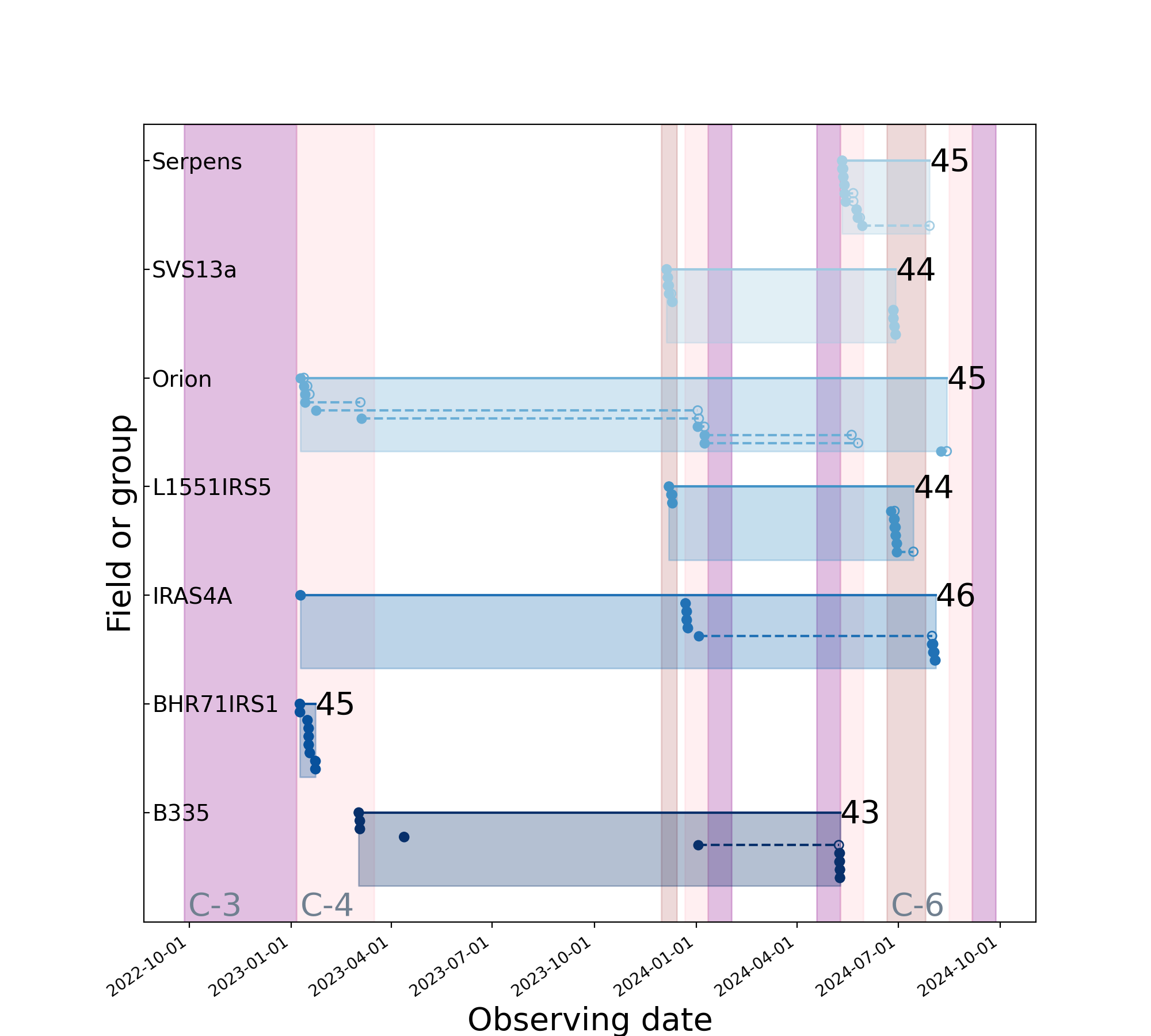}
\caption{Timeline of observations for each source or source group (``Orion'' represents V883-Ori, HOPS 373-B, and HOPS 108; ``Serpens'' represents Serpens emb 11 W, Serpens South-MM18a, and Serpens SMM1a).  Shaded blue bands represent the entire time between the first and last executions of any SB for a given source group on the y-axis.  Filled circles represent the start of each Scheduling Block.  In the cases for which an SB required multiple executions that were not performed on the same date, then an open circle indicates the final execution for a given SB.  In those cases, the dashed horizontal lines indicate the entire period during which an SB was started but incomplete; in some cases this was due to waiting for the array to return to an appropriate configuration. The numerical label next to the right-most dot for each source indicates the average number of antennas for executions of that group.  Pink shaded vertical bands show periods when the different `nominal' configurations (C-3, C-4, C-6) were available, with one of each labeled for guidance.} \label{fig:obstimeline}
\end{minipage}
\end{figure}

In this section, we report the pertinent details of the observations that were made with ALMA as part of the COMPASS LP \citep[proposal id: 2022.1.00316.L, see also][]{jorgensen2026}\footnote{\label{note1}\href{https://erda.ku.dk/vgrid/COMPASS/}{https://erda.ku.dk/vgrid/COMPASS/}}, specifically to provide the reader with enough context to accurately interpret the datasets that we present in Sect.~\ref{sec:dataproducts}.  We refer the reader to \citet{jorgensen2026} for more on the scientific justification of the setup.  Eleven sources were observed (see Table \ref{tab:imaging2}), each with nine tunings, and four adjacent spectral windows (SPWs) per tuning.  In Cycle 9, only one source (BHR71-IRS1) was completely observed with all frequency tunings.  Five other sources had between one and four tunings observed.  During Cycle 10, the remaining observations were queued and completed.  A graphical representation of the observations is given in Fig.~\ref{fig:obstimeline}.

Table \ref{tab:obspipelinelog}\ documents relevant observational information for each scheduling block (SB).  Where sources are in close proximity on the sky, they were clustered to be observed within the same SB to increase observing efficiency by decreasing calibration overheads; this was done specifically for the sources in Orion (V883-Ori, HOPS 373-B, and HOPS 108), and for the sources in Serpens or Serpens South (Serpens emb 11 W, Serpens South-MM18a and Serpens SMM1a).  In those cases, it should be noted that although the SBs differ in terms of which source name is included in the SB name, the targets are grouped together consistently (for example, for the nine SBs related to the Serpens sources, two begin with the source label ``Ser\_emb,'' five with ``SerpS-MM,'' and two with ``Serp\_SMM''). 

Table \ref{tab:spwinfo} (with the full table available at the CDS) is intended as an aid to understand frequency ranges, beam sizes, and noise levels for each tuning and SPW for a given SB. Each spectral window of 937.5 MHz in width has 3840 channels with a channel spacing of 244 kHz; the effective frequency resolution is 488.281 kHz, equivalent to a velocity resolution of 0.47–0.52 \kms\ across the targeted spectral range.  It is also apparent in Tables \ref{tab:imaging2} and \ref{tab:spwinfo} that beam sizes obtained were $\sim0.3-0.5^{\prime\prime}$ for most tunings of most sources, and $\sim0.15-0.2^{\prime\prime}$ for the two sources NGC1333-SVS13a and L1551-IRS5, which are each known to be tight binaries.  In Sect.~\ref{sec:imaging}, we describe the respective beam sizes in greater detail.

In this paper, we label the tunings numerically as 1-9, although it should be noted that the SBs generated by ALMA were differentiated by the source name and the letters a-i (e.g.,\ B335\_a\_07\_TM1 has the letter ``a'' following the source name ``B335'' and an underscore); our labels 1-9 correspond to progressively increasing frequencies (see Table \ref{tab:spwinfo}) and are consistent across all sources, whereas the lettered labels of a-i did not have a one-to-one correspondence to frequency range among all the SBs.  

Our project was set up with the ``system-defined calibration option,'' which automatically assigns appropriate calibrators from the ALMA Calibrator Source Catalog. The sources observed as phase, bandpass, and amplitude calibrators during each execution are listed in their entirety in Table \ref{tab:obspipelinelog}.  

It may be especially useful for archival researchers to note that one SB (HOPS-108\_b\_07\_TM1, in MOUS uid://A001/X2d20/X2fac) was reobserved when it was determined that the spectral setup was previously incorrect due to a technical issue, which resulted in non-continuous frequency coverage with neighboring tunings.  The correct SB is named HOPS-108\_bx\_07\_TM1 (in MOUS uid://A001/X3734/X73), and corresponds to the spectral setting 4 for the sources HOPS 108, HOPS 373-B, V883-Ori.  For another SB, NGC1333\_e\_07\_TM1, which corresponds to setting 5 for IRAS4A, one of the two executions seemed to have a truncated frequency range.  Fortunately, the conditions at the time of observations were good, such that the sensitivity of the single execution with the correct frequency range is sufficient for our study; therefore, we omit the execution with the inconsistent frequency coverage.  Other than the few issues mentioned here, affecting only about $2\%$ of the observations, the vast majority of observations were completed successfully.

\section{Data processing} \label{sec:dataproc}

\begin{figure*}[hbt]
\includegraphics[width=0.75\textwidth]{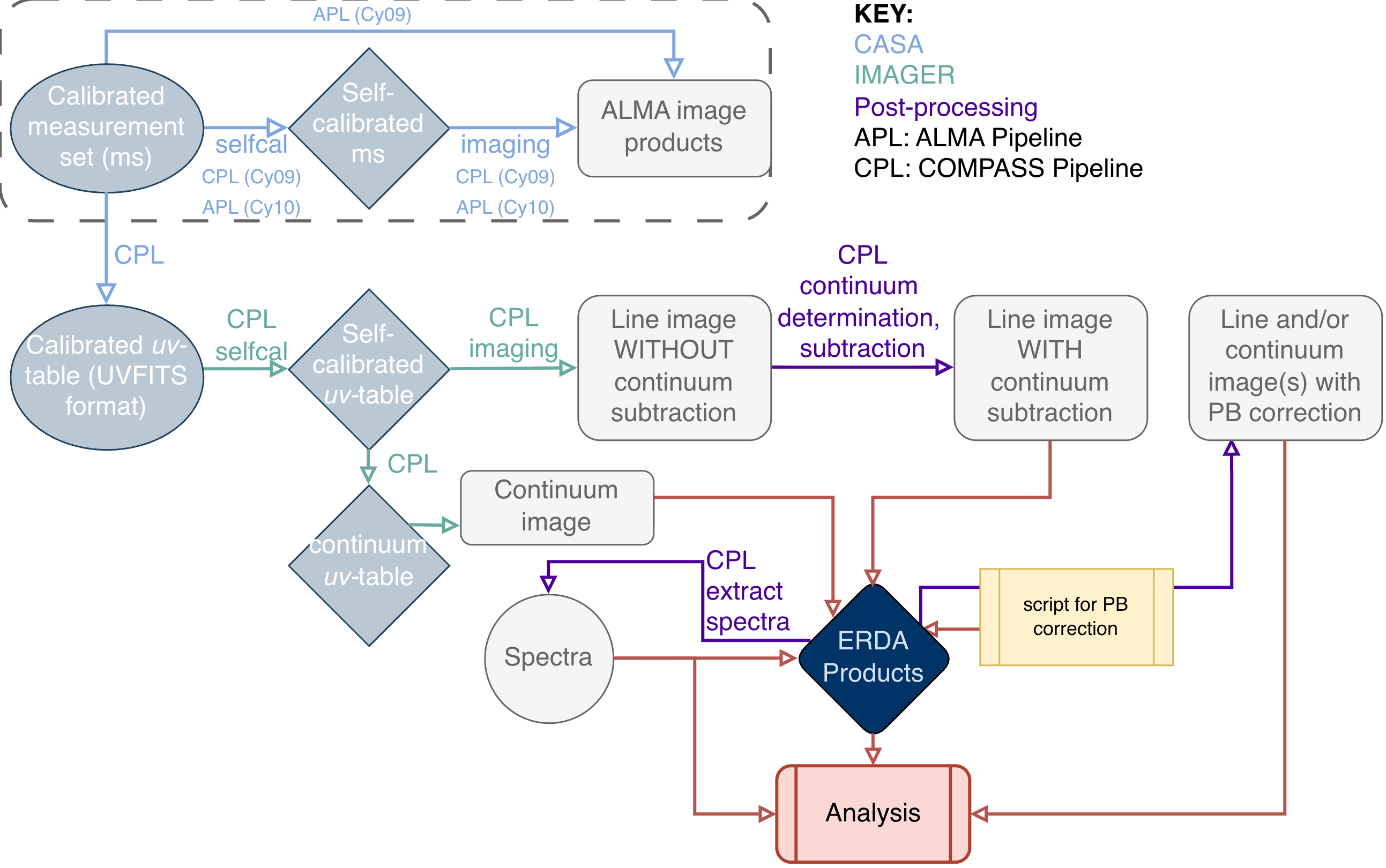}
\caption{Primary data flow, which uses \textsc{CASA} for calibration steps, and IMAGER for self-calibration and imaging.  We use CPL and APL to distinguish between steps of the COMPASS and ALMA Pipelines, respectively.  Dashed-outlined region indicates products that are delivered by ALMA, following the APL, and indicated for reference (see also Fig.~\ref{fig:casaflow}).  ``ERDA'' is the Electronic Research Data Archive at University of Copenhagen, and our team's primary data repository.
}\label{fig:imagerflow}
\end{figure*}

In this section, we explain the principal aspects of COMPASS data processing, which we refer to as the COMPASS Pipeline or ``CPL.'' Our data processing strategy aimed to meet several primary objectives: (1) accuracy; (2) reproducibility; (3) efficiency.  Our team utilized a combination of both IMAGER\footnote{ \href{https://imager.oasu.u-bordeaux.fr}{https://imager.oasu.u-bordeaux.fr}} and CASA \citep{CASA2022} for data processing, and in this section we describe the primary data flow utilized for scientific analysis in subsequent publications.  A diagram of the complete data flow is given in Figure \ref{fig:imagerflow}.  We deployed the CPL on the Dahu cluster at GRICAD\footnote{\href{https://gricad.univ-grenoble-alpes.fr}{https://gricad.univ-grenoble-alpes.fr}}, utilizing additional scheduler scripts to manage processing.  In Appendix \ref{sec:AAcasa} (see also Fig. \ref{fig:casaflow}), we present an analog data flow that utilized \textsc{CASA} exclusively (referred to as ``CPL-C''), in order to compare the imaging results of IMAGER with those of CASA (Appendix \ref{sec:imagervscasa}). 

The CPL is available on GitHub\footnote{\url{https://github.com/compass-alma-large-program/compass-data}}, and version 1.0, which was used to produce the data presented in the COMPASS first papers, is archived on Zenodo \citep{Maret25}. Data products produced by our team will be available through the ALMA Science Archive, following ALMA data standards for Large Programs.  Our team website\textsuperscript{\ref{note1}} also provides additional resources for data inspection. 

\subsection{Initial inspection and calibration}\label{sec:calibration}

The calibrated measurement sets were generated upon request to the European ALMA Regional Center (ARC), following the standard ALMA calibration pipeline which uses \textsc{CASA} \citep{CASA2022,Hunter2023}. Data delivered in Cycle 9 (observations before October 2023) were calibrated with ALMA Pipeline version 2022.2.0.64 (\textsc{CASA} 6.4.1.12), and the process did not include self-calibration.  Most of the data were delivered in Cycle 10 (observations after October 2023) and were calibrated with pipeline version 2023.1.0.124 (\textsc{CASA} 6.5.4.9), with the primary difference being that self-calibration using the task \textsc{hif\_selfcal} was incorporated (see Sect.~\ref{sec:selfcal} on self-calibration).  The COMPASS Pipeline (Figure \ref {fig:imagerflow}) uses as input the calibrated measurement sets (specifically, the column that corresponds to data without self-calibration applied), which are extracted to $uv$\ table format (as ``UVFITS'' file type) in IMAGER.  

Upon delivery of each data package, our team manually checked the weblogs to note any major issues during observations or calibration, and took an initial look at the images for each spectral window within a given tuning.  The only observational issues encountered were the two cases described in Sect.~\ref{sec:observations}, and the vast majority of the bandwidth for our project was observed successfully.  We show in Table \ref{tab:obspipelinelog} that common calibrators were used across most tunings for a given target, which lends consistency and robustness to phase, bandpass, and amplitude calibrations.   

\subsection{Self-calibration}\label{sec:selfcal}

The CPL includes a step of self-calibration using IMAGER. The self-calibration was performed on the line + continuum $uv$\ table. To achieve this, all spectral channels were averaged together to improve the signal-to-noise ratio. The self-calibration solution was derived from this averaged table, using three iterations with an integration time of 10 second each. Finally, the resulting phase corrections were applied to all individual channels to produce a self-calibrated line + continuum $uv$\ table. We found that this increased the dynamic range of the signal in the continuum images by up to an order of magnitude, while decreasing the noise by a comparable factor. We also tried to apply self-calibration to the amplitude after the phase, but the data quality did not improve significantly. Therefore, we decided to apply the self-calibration only to the phase.  A more detailed description of the continuum image quality improvement due to self-calibration is found in Appendix \ref{sec:impr-self-calibr}.

Although the self-calibration improves the continuum images significantly in some cases, it has a much smaller effect on the lines. The comparative benefit for continuum is due to some sources having large continuum fluxes.  Without self-calibration, the continuum images are limited by the dynamic range, whereas the dynamic range improves with the self-calibration step, especially for the sources with the brightest emission (see Appendix \ref{sec:impr-self-calibr}). On the other hand, the line fluxes are in many cases relatively weaker compared with the continuum, and the spectral line image cubes are noise-limited.\footnote{In practice, the thermal noise is computed by IMAGER from the visibility weights, and the cubes have noise similar to the thermal level.  We also used the ALMA sensitivity calculator to estimate the thermal noise level for some representative line image cubes, taking into account corresponding integration time on source (the median on-source integration time is 40 minutes for this project), number of antennas, and typical conditions, for which we determine expected noise levels within a factor of $\sim2$\ of the mean RMS values in Table \ref{tab:imaging2}.}  Whereas the marginal gains were more significant for continuum images than line data cubes, we applied self-calibration to the line data as well, in order to have a consistent calibration between the continuum and the lines.

Although the CPL incorporates self-calibration in IMAGER, we reiterate for thoroughness that some data (specifically from observations during or after December 2023) were processed with a version of the ALMA Pipeline that also incorporated self-calibration. On the other hand, the observations toward BHR71IRS1 (and a few others) were delivered in early 2023 without self-calibration; this is when our team was developing the CPL, and we decided to integrate all further steps, including self-calibration, in IMAGER.  Thus, the CPL takes as input the calibrated (without self-calibration) data.  For testing purposes, we ran self-calibration in CASA with the \textsc{hif\_selfcal} task in \textsc{CASA} 6.5.4 for any data that were delivered without it, as indicated in Fig.~\ref{fig:imagerflow}.\footnote{With \textsc{CASA} 6.5.4, an automated self-calibration can be run, which stops when either there is no improvement in the $S/N$ or the synthesized beam has significantly changed \citep[Tobin et al.~in prep, see also a thorough description of an earlier implementation in][]{Ohashi2023}.  The results have been robustly checked against manual self-calibration.} No significant difference was determined between measurement sets that were self-calibrated with \textsc{IMAGER} compared with \textsc{CASA}.

\subsection{Imaging} \label{sec:imaging}
\begin{table*}[ht!]
\caption{Imaging parameters.}
\label{tab:imaging2}
\centering
\footnotesize
\begin{tabular}{lccccccc}
\hline\hline
Target & \multicolumn{2}{c}{Phase center} & Line HPBW\tablefootmark{(a)} & $\overline{\sigma_{\textrm{cube}}}$\tablefootmark{(b)} & Continuum HPBW\tablefootmark{(a)} & $\overline{\sigma_{\textrm{cont}}}$\tablefootmark{(b)} \\
\cline{2-3} 
 & RA (ICRS) & Dec (ICRS) & min--max ($^{\prime\prime}$) & (mJy beam$^{-1}$ ch$^{-1}$)& min--max ($^{\prime\prime}$) & (mJy beam$^{-1}$) \\
\hline
NGC~1333-SVS13A & 03:29:03.756 & $+$31:16:03.80 & 0.14--0.18 & 2.20 & 0.14--0.17 & 0.14 \\
NGC~1333-IRAS4A & 03:29:10.432 & $+$31:13:32.12 & 0.26--0.48 & 2.51 & 0.26--0.48 & 0.32 \\
L1551-IRS5 & 04:31:34.160 & $+$18:08:04.72 & 0.13--0.16 & 2.19 & 0.12--0.16 & 0.15 \\
HOPS 108 & 05:35:27.084 & $-$05:10:00.07 & 0.28--0.41 & 2.64 & 0.28--0.40 & 0.25 \\
V883-Ori & 05:38:18.100 & $-$07:02:26.00 & 0.28--0.41 & 2.64 & 0.27--0.43 & 0.18 \\
HOPS 373-B & 05:46:30.905 & $-$00:02:35.20 & 0.28--0.42 & 2.56 & 0.27--0.41 & 0.13 \\
BHR71-IRS1 & 12:01:36.499 & $-$65:08:49.38 & 0.36--0.44 & 2.17 & 0.36--0.44 & 0.22 \\
Serpens emb 11 W & 18:29:06.720 & $+$00:30:34.30 & 0.26--0.44 & 2.64 & 0.26--0.42 & 0.10 \\
Serpens SMM1a & 18:29:49.793 & $+$01:15:20.20 & 0.25--0.43 & 2.67 & 0.25--0.42 & 0.32 \\
Serpens South-MM18a & 18:30:04.118 & $-$02:03:02.55 & 0.24--0.42 & 2.69 & 0.24--0.41 & 0.22 \\
B335 IRS & 19:37:00.894 & $+$07:34:09.59 & 0.23--0.49 & 2.15 & 0.23--0.48 & 0.14 \\
\hline
\end{tabular}
\tablefoot{\tablefoottext{a}{For a given target, minimum and maximum values of geometric mean beam sizes (half-power beam width, or HPBW) are given for line cubes and continuum images, as noted (see also Fig. \ref{fig:continuumintensityandbeam}).}\tablefoottext{b}{ $\overline{\sigma}$ is the average RMS noise in the data (line cubes or continuum images, in respective columns) for a given source; in the case of line cube, $\overline{\sigma_{cube}}$\ is measured per 0.47-0.52 km s$^{-1}$\ channel}. Respective values for all spectral line cubes are given in Table \ref{tab:spwinfo}.}
\end{table*}

\begin{table*}
\centering
\caption{Continuum properties. \label{tab:continuum}}
\begin{tabular}{lccccc}
\hline\hline
Target\tablefootmark{(a)} & \multicolumn{2}{c}{Continuum peak position\tablefootmark{(b)}} && \multicolumn{2}{c}{Fit to continuum flux\tablefootmark{(c)}} \\
\cline{2-3}\cline{5-6}
(abbr.) & RA (ICRS) & Dec (ICRS) && $F_{{\rm int, 300 GHz}}$ (Jy) & $\alpha$\\
\hline
SVS13a              & 03:29:03.749 & +31:16:03.72 && $0.52\pm 0.02$   & $2.8\pm 1.0$ \\
IRAS4A              & 03:29:10.438 & +31:13:32.03 && $0.63\pm 0.02$   & $2.5\pm 1.0$ \\
L1551IRS5           & 04:31:34.166 & +18:08:04.59 && $0.97\pm 0.04$   & $2.4\pm 1.0$ \\
HOPS108             & 05:35:27.085 & -05:10:00.08 && $0.041\pm 0.002$ & $2.6\pm 1.0$ \\
V883Ori             & 05:38:18.100 & -07:02:25.99 && $0.72\pm 0.03$   & $3.0\pm 1.0$ \\
HOPS373             & 05:46:30.905 & -00:02:35.18 && $0.128\pm 0.005$ & $3.0\pm 1.0$ \\
BHR71               & 12:01:36.475 & -65:08:49.36 && $0.82 \pm 0.03$  & $3.1\pm 1.0$ \\
Emb11W              & 18:29:06.625 & +00:30:33.89 && $0.097\pm 0.004$ & $3.1\pm 1.0$ \\
SMM1a               & 18:29:49.804 & +01:15:20.34 && $1.40\pm 0.05$   & $2.1\pm 1.0$ \\
MM18a               & 18:30:04.123 & -02:03:02.64 && $0.35\pm 0.01$   & $2.3\pm 1.0$ \\
B335                & 19:37:00.903 & +07:34:09.41 && $0.077\pm 0.003$ & $2.6\pm 1.0$ \\
\hline
\end{tabular}
\tablefoot{\tablefoottext{a}{Targets are referred to with their abbreviated names, consistent with \citet{jorgensen2026}.} \tablefoottext{b}{The continuum peak position corresponds to the continuum source associated with the primary COM emission in each field. The coordinates represent the median coordinates obtained from two-dimensional Gaussian fits to the source in the individual continuum images of all thirty-six spectral windows.} \tablefoottext{c}{Integrated flux at 300 GHz, $F_{{\rm int, 300 GHz}}$, and the spectral index, $\alpha$, are determined from a power-law fit to the continuum fluxes across the observed spectral range (see Sect.~\ref{sec:quality}, and Fig.~\ref{fig:continuumflux}) of the form $F = F_{{\rm int, 300 GHz}}(\nu/300\rm{\ GHz})^\alpha$.}}
\end{table*}

\begin{figure*}[!htb]
\includegraphics[width=\textwidth]{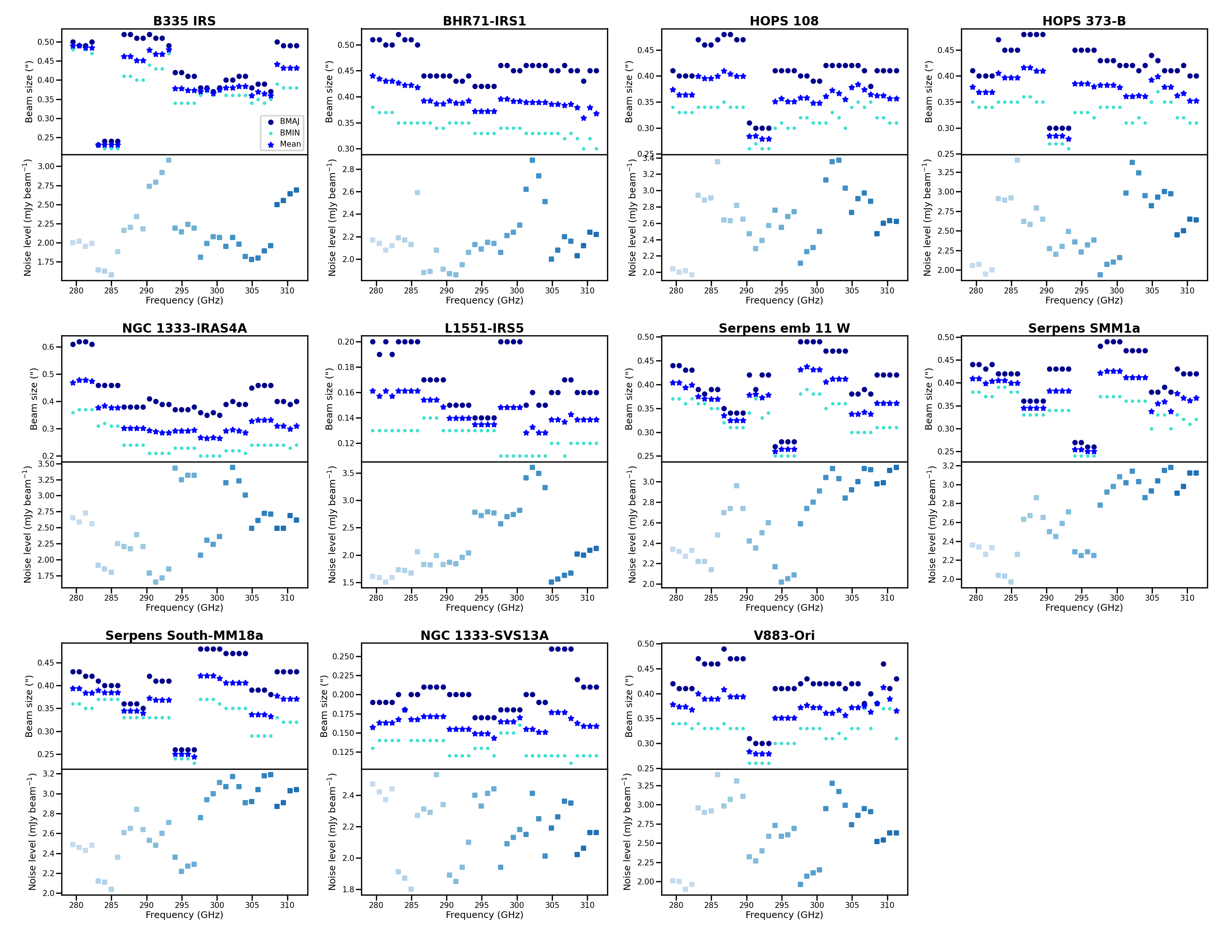}
\caption{Beam size (upper part of each panel) and RMS noise level per channel (lower part of each panel) in line cubes as a function of frequency for all sources in the sample (see also Table \ref{tab:spwinfo}).  For each source, the upper panel shows the major and minor axes of the beam ($\theta_{\textrm{maj}}, \theta_{\textrm{min}}$) for each independent spectral cube, as well as the ``geometric mean,'' defined as $\sqrt{(\theta_{\textrm{maj}})(\theta_{\textrm{min}})}$.  Symbols in the lower panel get progressively darker in groups of four, representing the four spectral windows in each tuning.  The requested noise level for the observations was equivalent to 3.5 mJy beam$^{-1}$ per channel; the line cubes for all sources and tunings meet or exceed this quality standard.} \label{fig:continuumintensityandbeam}
\end{figure*}

Imaging was performed separately for each source, frequency tuning, and spectral window.  The CPL utilizes IMAGER for the imaging and deconvolution steps.  The two tasks are \textsc{uv\_map} and \textsc{clean}.  In the first step, the dirty map and beam are generated from the $uv$-data.  Each image is a square with side length defined as twice the size of the primary beam; the sensitivity drops significantly beyond this region.  Each tuning for a given source is ultimately imaged with a uniform pixel size for that source; to determine the best uniform value, we select the smallest pixel size and the largest image size determined automatically by IMAGER among all the tunings for that source.  In the latter step (\textsc{clean}), the dirty map and beam are deconvolved using the \textsc{Hogbom CLEAN} algorithm to produce the clean and residual images.  A robust\footnote{Robust parameter values in IMAGER can range from 0 to infinity, which is different from the convention invoked in CASA's \textsc{tclean}.} value of 1 is used, which in IMAGER means that cells with natural weighting above this threshold will be given the (uniform) weight of 1.  Our selected value balances sensitivity and angular resolution.  We show these values graphically in Fig.~ \ref{fig:continuumintensityandbeam}, and a complete list of output image parameters, including beam sizes and RMS noise, is given in Table \ref{tab:spwinfo} (and online).  As shown in the top portions of each panel in Fig.~ \ref{fig:continuumintensityandbeam}, the beam sizes for each tuning meet the requested criteria of $\sim0.3-0.5^{\prime\prime}$ for most sources, and $\sim0.15-0.2^{\prime\prime}$ for the two sources NGC1333-SVS13a and L1551-IRS5. In only a handful of cases among all the image cubes, the resulting half-power beam width (HPBW) are slightly smaller than the requested range; these also correspond to observations with configurations that had slightly longer baselines compared with those for other tunings for a given source (see Table \ref{tab:obspipelinelog} for observing dates and corresponding maximum baselines). While the deviations are minimal, analysis should always take into account the actual beam size. A summary of imaging parameters for each source and each data set (line cubes and continuum images) is given in Table \ref{tab:imaging2}.  The noise ($\sigma_{cube}$) for each image cube is determined using IMAGER's \textsc{statistics} command, which computes the RMS outside the inner quadrant of the cleaned image; the noise ($\sigma_{cont}$) for each continuum image in Table \ref{tab:imaging2} is determined by astropy.stats  sigma\_clipped\_stats. Given that the observing proposal specified a RMS noise of 2.5 mJy beam$^{-1}$ over a bandwidth of 1 km s$^{-1}$, corresponding to 3.5 mJy beam$^{-1}$ per channel, the observations exceeded the nominal standard. 

We produce images without primary beam correction, as it is necessary to keep the noise uniform for the later step of continuum identification and subtraction (Sect.~\ref{sec:continuumid}).  We generate cubes with no truncation at a specific primary beam level (within a defined field-of-view), allowing the possibility to study extended emission and/or neighboring sources at the edge of the nominal field of view, as in the case of BHR71-IRS1 and BHR71-IRS2. In addition, we provide a script that will allow a user to generate their own cubes and spectra that have been corrected by the primary beam response. 

Imaging interferometry data with many pixels and channels can be a computationally expensive and time-intensive process.  Adding complexity, we had to test and optimize the continuum subtraction method for our datasets (see Sect.~\ref{sec:continuumid}), which required an iterative approach of imaging, inspection, and scientific validation.  Based on our evaluation of imaging with both IMAGER and CASA (both of which used the Hogbom CLEAN algorithm), we identified IMAGER as significantly faster and therefore suitable for the case of reimaging.  IMAGER is optimized to perform operations using internal buffers, which avoids as much as possible saving intermediate data on disk, and therefore speeds up the process compared to \textsc{tclean} with CASA.  We benchmarked our CPL against one done completely with CASA for the case of BHR71-IRS1, as described in Appendix \ref{sec:AAcasa}. A full description of the comparison between images generated with IMAGER and CASA is given in Appendix \ref{sec:imagervscasa}.

\subsection{Continuum determination and subtraction} \label{sec:continuumid}

\begin{figure*}[!hbt]
\includegraphics[width=0.45\textwidth]{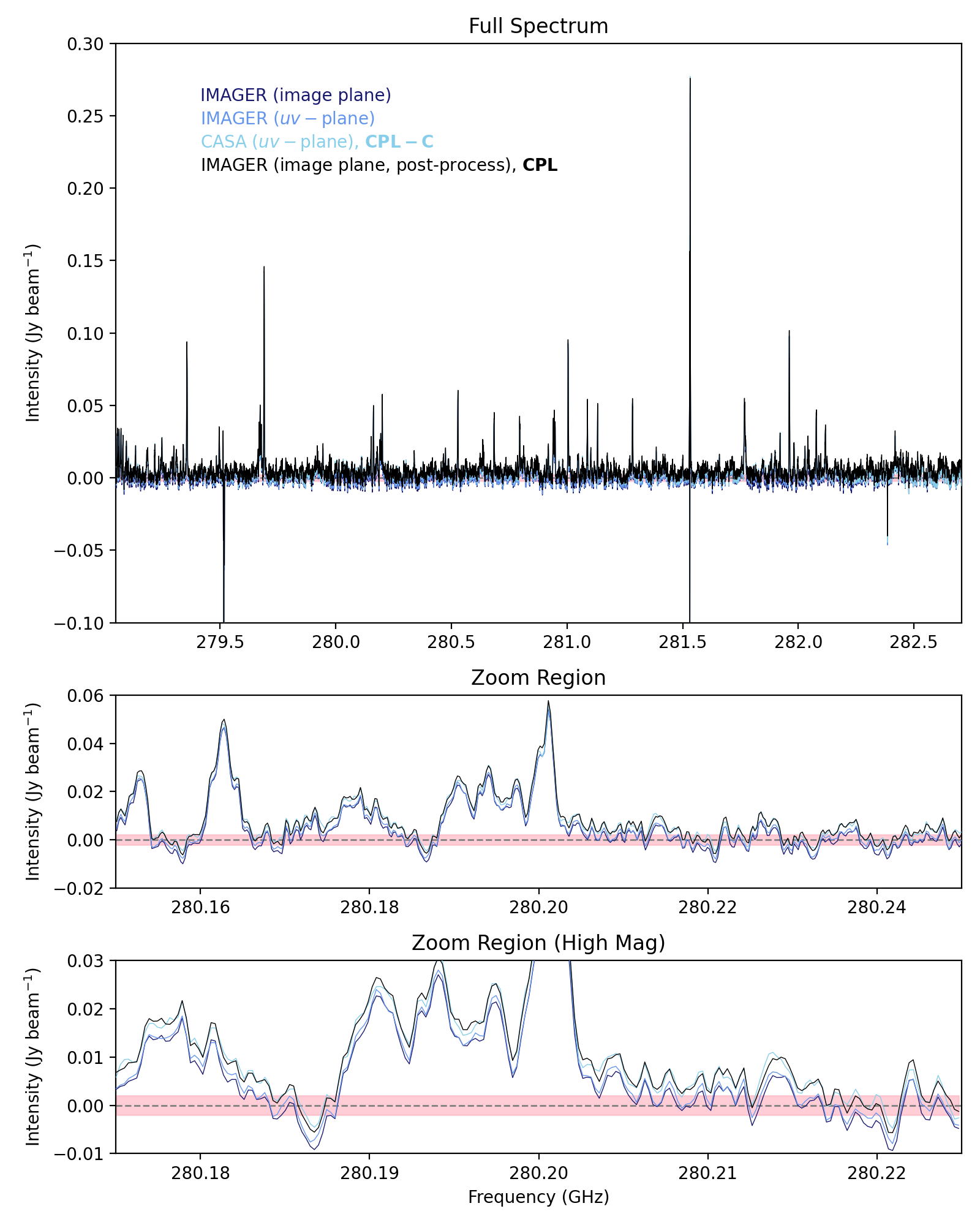}
\includegraphics[width=0.45\textwidth]{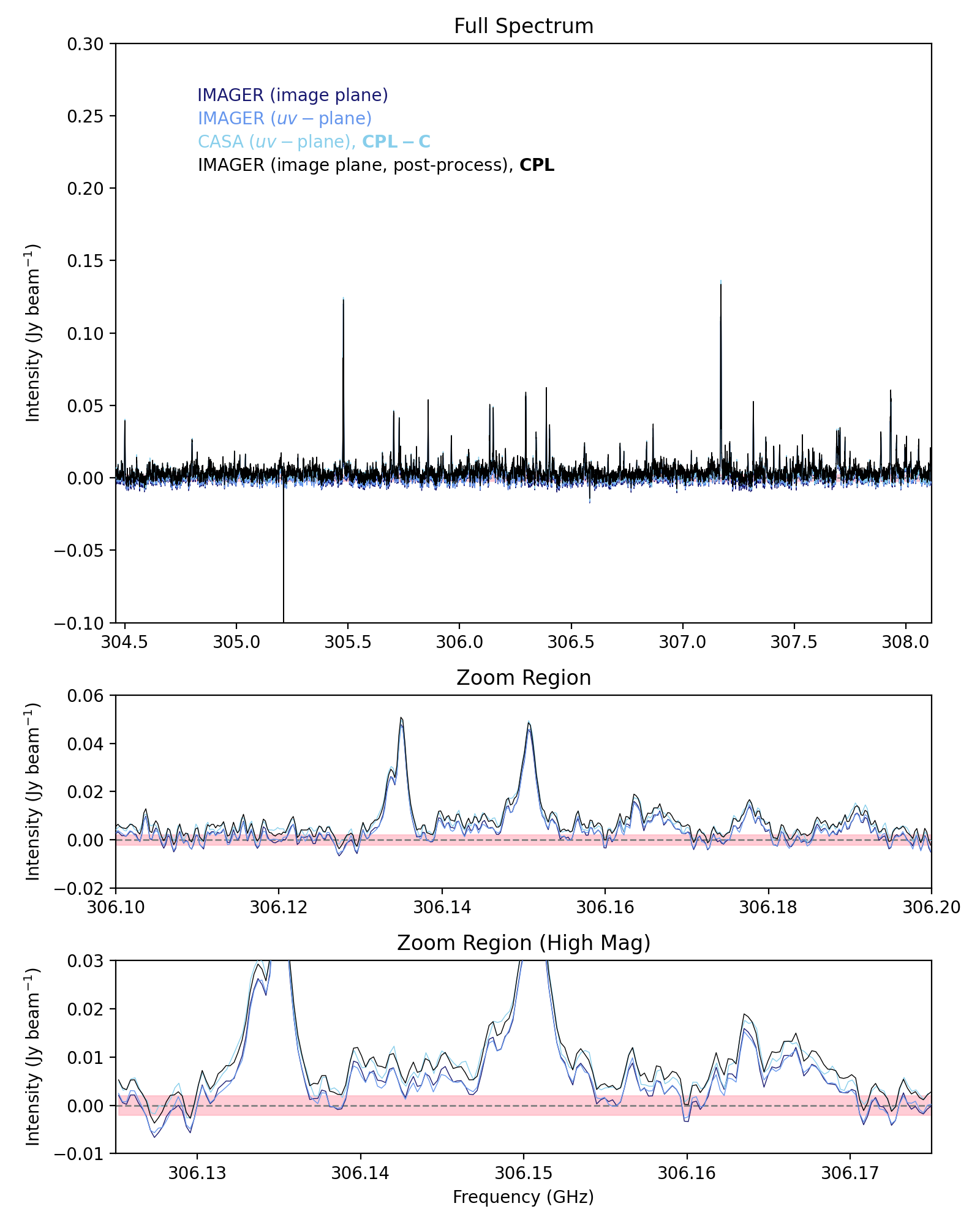}
\caption{Spectra for BHR71-IRS1 tunings 1 (left) and 8 (right), showing (with different colors) a comparison of the main continuum identification and subtraction methods described in this paper (see also Appendix \ref{sec:continuumalternatives}), specifically post-processing in the image plane \citep[the primary CPL method for spectral line cubes, see also][]{jorgensen2016}, using CASA and performing continuum subtraction in the \textit{uv}-plane (CPL-C, see Appendix \ref{sec:AAcasa}), and using IMAGER with continuum subtraction in the \textit{uv}-plane and image plane.  Spectra were extracted from the phase center for BHR71-IRS1.  Upper panels show the full spectra spanning the entire spectral coverage for a given setting.  Middle and lower panels show progressive zoomed-in regions to show more spectral detail.  Horizontal dashed line marks 0 Jy bm$^{-1}$, and the pink shaded band shows the $\pm1\sigma$\ RMS noise level (where $\sigma\sim2$\ mJy bm$^{-1}$ per channel, see Table~\ref{tab:imaging2}).} \label{fig:contsub}
\end{figure*}

Precise determination of the continuum level in a spectral line cube is crucial for deriving reliable continuum-subtracted spectra and constructing optimized continuum images, but it can be particularly challenging in line-rich spectral surveys.  Broadly, we can distinguish two categories of methods for continuum identification and subtraction: those that statistically determine the continuum and the line contributions; and those that depend on identifying certain channels that have no contribution from spectral lines.  Datasets such as ours with a large number of spectral emission lines and high line density can pose a challenge for identifying channels without significant line emission that could be designated as continuum.  Therefore, we opted to incorporate a statistical method applied with post-processing in the image plane, and we show a comparison of example spectra in Fig.~\ref{fig:contsub}.  Readers can refer to Appendix \ref{sec:continuumalternatives} for descriptions of additional methods that we investigated; in Fig.~\ref{fig:contsub} it is shown that although small differences exist, the CPL and CPL-C (which uses exclusively CASA, see Appendix \ref{sec:AAcasa}) datasets provide scientifically equivalent results for most science cases. 

We primarily utilized the method of \citet{jorgensen2016}, which was previously applied to IRAS 16293-2422 in PILS.  This method has been shown to be particularly valuable in cases where there is a very limited ($\leq$ 10\%) bandwidth that is line-free.  While more information can be found in \citet{jorgensen2016}, briefly, the method is able to statistically determine the continuum and line contribution for the spectrum associated with each pixel in the image cube.  In the case of a pixel in which only continuum emission and/or noise is present, the density distribution of flux values in that pixel can be fitted with  a symmetric Gaussian centered on the continuum level, and its width corresponds to the channel-by-channel rms noise.  On the other hand, for a pixel which has line emission present, the Gaussian is modified by an exponential tail toward higher values.  The estimated per-pixel continuum level can then be subtracted from each pixel in the image, and this is done for each spectral window. 

Our chosen method for spectral cube processing operated in the image plane.  It is also possible to do the continuum identification in the $uv$-plane, which tends to lead to a nearly optimal continuum image. The continuum identification in the $uv$-plane was done with the task \texttt{uv\_preview} in IMAGER. This task computes spectra toward the phase center at different spatial scales, using three different tapers. In order to be sensitive to emission lines of different widths, each of these spectra is then smoothed by averaging over a different number of channels (by default 4 or 16). Finally, an iterative scheme is used, blanking out ranges of channels above 3$\sigma$ (the default value, although other clipping values were also tested), to converge toward a Gaussian histogram of the intensity distribution of all channels, which nominally represents the noise distribution around the continuum level.  This method generates a (mostly) line-free visibility table with a large bandwidth, suitable for continuum imaging. Since resolution depends on frequency, the larger bandwidth enhances $uv$ coverage\footnote{This is often referred to as bandwidth synthesis.}, enabling the creation of images with reduced noise compared to those for which continuum is derived from the image plane. In practice, continuum images produced from data cubes in the image plane are typically $2-3$ times noisier than those generated in the $uv$-plane. 

However, continuum-subtracted line cubes generated with this method (or other comparable methods) were more likely to display over-subtraction in some spectral ranges, compared to the statistical method operating in the image-plane, as described above (see Fig.~\ref{fig:contsub}).   Therefore, we adopted a ``dual'' approach: line data cubes were created by imaging the line-plus-continuum visibility tables and then identifying and subtracting the continuum in the image plane using the statistical method of \citet{jorgensen2016}, while continuum images were generated by identifying line-free channels directly in the $uv$-plane and imaging them as continuum.  For each field,  Table \ref{tab:continuum} gives the coordinates of the peak continuum emission for each field. Table \ref{tab:continuum} also gives the continuum flux density, $F_{{\rm int, 300 GHz}}$, at a representative frequency (300 GHz) for each source (see also Sect.~\ref{sec:quality}).

\subsection{Data products} \label{sec:dataproducts}
In this section, we outline the data products that have been generated as part of our project, and which will be made publicly available. For the context of this discussion, we refer to these as ``enhanced data products'' (EDPs). Figure \ref{fig:imagerflow} shows how each data product was generated and fits into the overall data flow. EDPs include:
\begin{itemize}
    \item Image cubes of each tuning and SPW for each source, generated with the CPL.  These images are provided \textit{without} continuum subtraction nor primary beam correction.  [Format: FITS]
    \item Image cubes as above, but these images are provided \textit{with} continuum subtraction and \textit{without} primary beam correction. [Format: FITS]
    \item Continuum images of each tuning and SPW for each source, \textit{without} primary beam correction.  [Format: FITS]
    \item Self-calibrated UV tables of each tuning for each source.  These data products would allow the reader to work with the IMAGER process of the CPL. [Format: UV tables]
    \item Spectra extracted from one or more selected lines of sight for each source. For all sources, a spectrum is extracted from the pixel corresponding to the location of the peak continuum intensity (Table~\ref{tab:imaging2}).  Additionally, in cases where offset spectra are relevant to the team's analysis, we provide the spectra.   [Format: ASCII]
    \item Python script to perform the primary beam correction of images as a post-processing step. [Format: Text file]
    \item Python script to extract spectra with the primary beam correction applied. [Format: Text file]
\end{itemize}

Additionally, we remind the reader that standard data products are made available by the ALMA Observatory.  These include the ``raw'' (uncalibrated) measurement sets in the native ALMA Science Data Model (ASDM) format, if one wishes to re-run the calibration, which can be downloaded from the ALMA Science Archive (ASA)\footnote{\href{https://almascience.org/aq/}{https://almascience.org/aq/}}.  As shown in Fig. \ref{fig:imagerflow} (also Fig. \ref{fig:casaflow}), the CPL begins with the calibrated measurement set, which can be requested from the ALMA Regional Centers.  Additionally, ALMA Pipeline imaging products for all targets and spectral windows are available from ASA.

\subsection{Quality control} 
\label{sec:quality}

\begin{figure*}[!htb]
\includegraphics[width=0.32\textwidth]{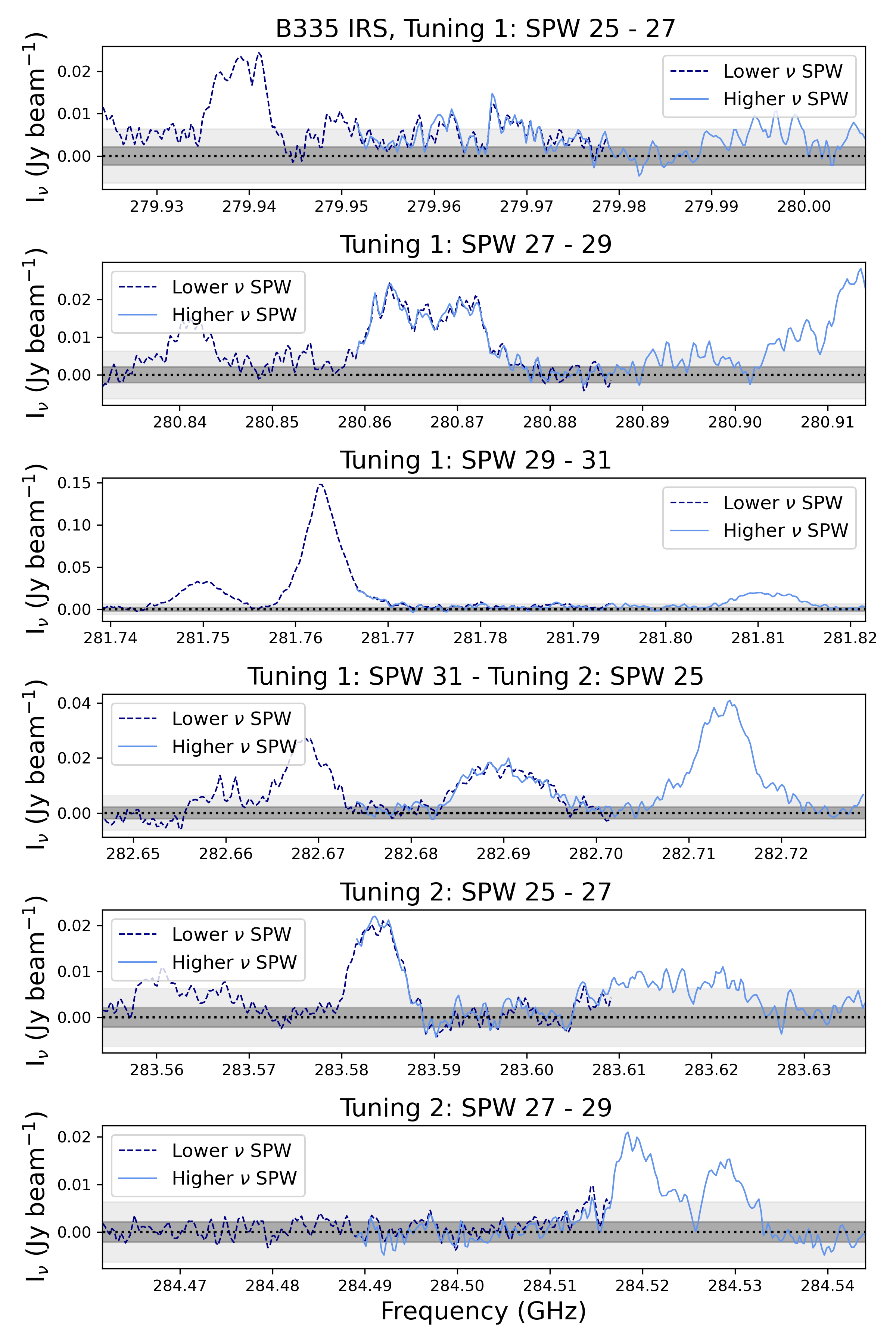}
\includegraphics[width=0.32\textwidth]{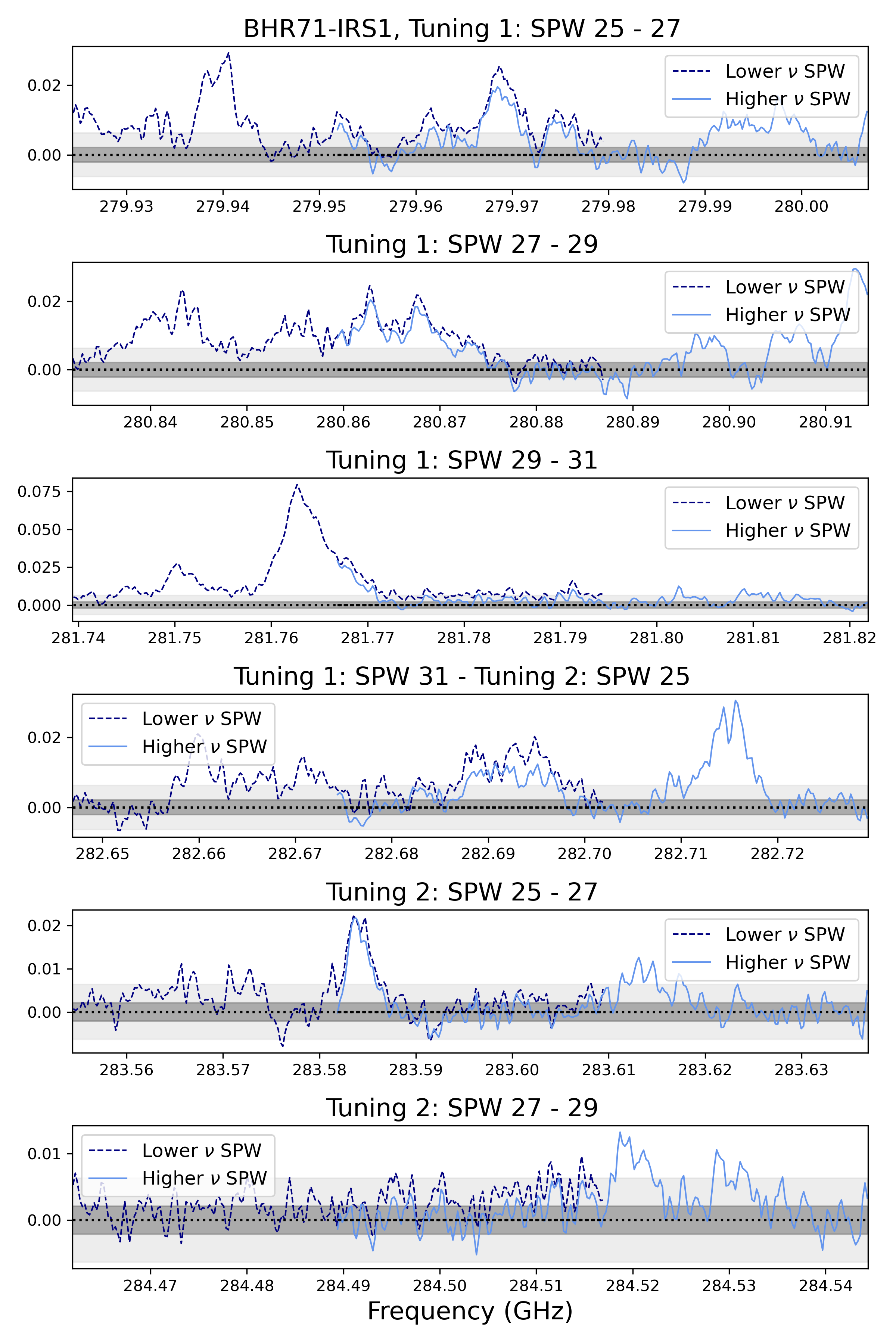}
\includegraphics[width=0.32\textwidth]{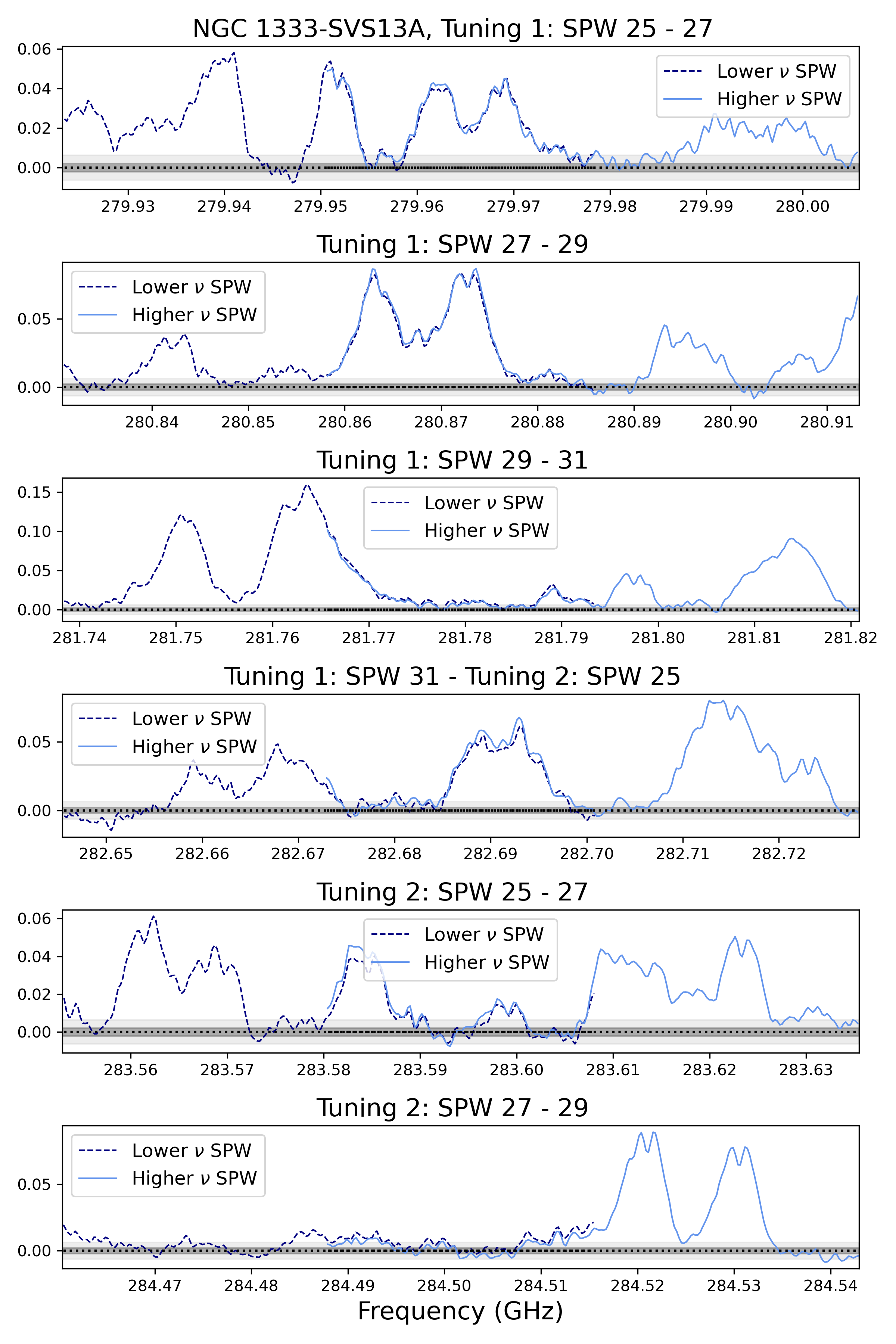}
\caption{Selected baseline-removed spectra spanning overlapping frequency regions between spectral windows (SPWs) in three sources: B335 IRS (left), BHR71-IRS1 (center), and NGC1333-SVS13A (right).  Spectra are obtained from the pixel at the continuum peak in Table \ref{tab:continuum}; units of Jy beam$^{-1}$\ correspond to the beam of the respective data cube (see Table \ref{tab:spwinfo}).  In each case, the two spectra are labeled as ``lower'' and ``higher'' frequency ($\nu$) SPWs, indicating the relative frequency range covered by each spectrum with respect to the other.  Dark and light gray shaded horizontal bands indicate $1\sigma$ (2 mJy bm$^{-1}$) and $3\sigma$ levels above and below a value of 0 (dotted horizontal line).  Here we demonstrate various scenarios of overlapping spectral ranges, including overlapping SPWs that fall within the same or different tunings, as well as frequency ranges with strong or weak spectral features. Selected SPWs are the same for each of the three sources shown here in order to demonstrate comparable spectral lines among panels in the same row.  } \label{fig:overlapregions}
\end{figure*}

In order to generate continuous spectra across a wider range than a single SPW and tuning, contiguous SPWs were set up to have 27.5 MHz of overlapping bandwidth, which we depict in Fig.~\ref{fig:overlapregions} for selected frequency ranges.  We inspected each overlapping range to verify that the continuum subtraction and flux levels were consistent. In Fig.~\ref{fig:overlapregions}, we show selected frequency ranges with and without strong emission features in an effort to demonstrate the relative effects of slight variations in flux levels between contiguous SPWs.  

\begin{figure*}[hbt]
\includegraphics[width=0.95\textwidth]{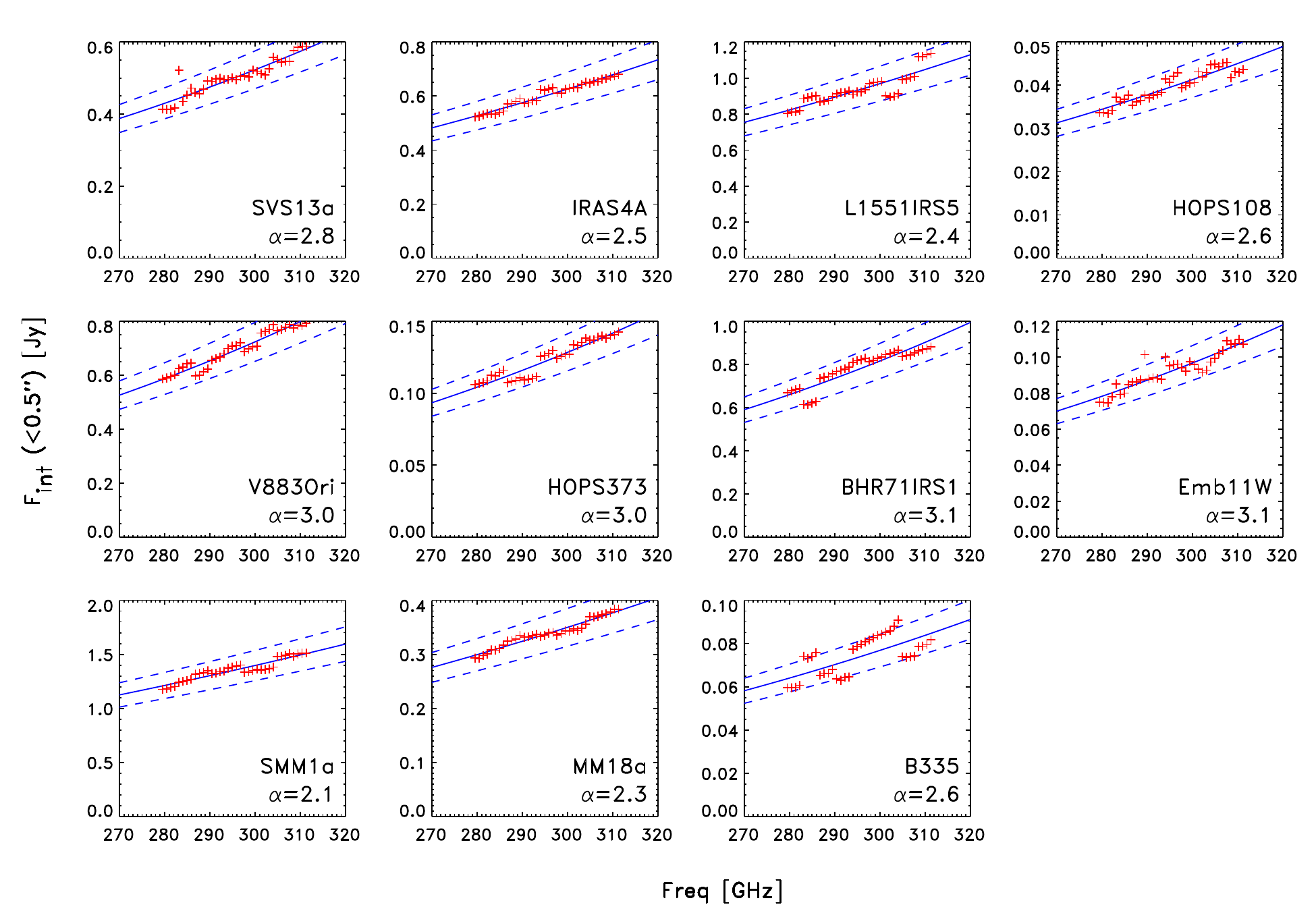}
\caption{Continuum flux density extracted from a $0.5\arcsec$ radius at the location of the peak continuum flux for each field, as indicated in Table \ref{tab:imaging2}.  The spectral index $\alpha$\ values are given in each panel, corresponding to the solid blue lines that indicate the best fits.  Blue dashed lines indicate flux uncertainty of 10\%.} \label{fig:continuumflux}
\end{figure*}

Differences are generally lower than the $1\sigma$ level, and in very few cases up to about the $3\sigma$ level.  It is apparent that the continuum level is slightly higher in a lower frequency SPW relative to the next higher frequency SPW.  This is to be expected, because each SPW with a bandwidth of 937.5 MHz should have a continuum level represented by a slight power law (see also Fig.~\ref{fig:continuumflux}), whereas in practice, a single continuum level is subtracted per pixel for all channels.  This ultimately leads to a slight over-subtraction in the lower frequency channels and a slight under-subtraction in the higher frequency channels of a given SPW.  The effect on studies of emission lines stronger than a few $\sigma$\ at the edge of a SPW should not be significant; studying weak or blended spectral features near the edge of a SPW should be handled with caution.  If a large enough offset across an entire SPW were to be detected between adjacent SPWs, one could be scaled relative to the other; however, we determine that the flux agreement from one SPW to the next is sufficiently good such that scaling would introduce a level of uncertainty comparable to or greater than the present uncertainty.

Additionally, we evaluated the continuum flux density across all SPWs observed for a given source.  Figure \ref{fig:continuumflux} shows the continuum flux density as a function of frequency for each SPW and all eleven sources in the sample; the continuum flux was extracted from a $0.5\arcsec$\ radius aperture at the location of the peak continuum emission associated with the principal COM emission in the primary-beam corrected image of each field \citep[Table \ref{tab:continuum}, see also][]{jorgensen2026}.  As a means of parameterizing the variation of the flux with frequency, we fit the measured continuum flux densities as a function of frequency for the individual sources with power-law functions of the form $F = F_{{\rm int, 300 GHz}}(\nu/300\rm{\ GHz})^\alpha$.  The derived spectral index values, $\alpha$, for these sources range from 2.1 to 3.1, with an uncertainty of $\pm1.0$ (see also Table \ref{tab:continuum}). Such spectral indices suggest that the emission stems from marginally optically thick dust components with an opacity law $\kappa_\nu \propto \nu^\beta$ with $\beta \sim 1$ \citep{prosacpaper}, where $\beta = \alpha - 2$.  Obtaining data over a broader frequency range would enable a reduction in the calculated error. On the other hand, the consistency we find such that the continuum fluxes across numerous settings follow a power-law (within the flux uncertainty of 10\% assumed for the observations), indicates that our calibration scheme is apparently robust. 

\section{Summary} \label{sec:conc}

This paper presents data-centric aspects of the COMPASS LP, with an aim to facilitate sufficient understanding of the data that form the basis of subsequent scientific analysis by the COMPASS collaboration \citep[see also][]{jorgensen2026,coutens2026}.  The scope of the paper includes the data processing once observations were completed, ultimately resulting in enhanced data products that we make available to the community.

In total, 11 embedded protostellar regions were targeted with nine spectral setups each with ALMA (covering the frequency range 279 to 312 GHz); each of the nine spectral setups comprised four spectral windows.  We described the data challenges associated with such datasets, and presented our methodology to address such challenges.

We used both CASA and IMAGER software for calibration and imaging, as diagrammed in Fig. \ref{fig:imagerflow}.  For imaging, the computational optimization of IMAGER was crucial for establishing the various imaging parameters and validating the procedure for all datasets.  Once the optimized method is determined, CASA could also be used for producing image products, as described in Appendix \ref{sec:AAcasa}.

Post-processing in the image plane following the method of \citet{jorgensen2016} was necessary for the continuum subtraction, as described in Sect.~\ref{sec:continuumid}, due to the high density of spectral line emission in the spectral axis.  We describe this in the context of continuum identification in the $uv$\ plane, which we used for generating continuum images.

The data products we provide are the most expansive, uniform dataset targeting the study of chemical composition of low-mass protostars to date.  Such a campaign was made possible via an ALMA Large Program, and provides an early glimpse into the possibilities of the ALMA Wideband Sensitivity Upgrade (WSU) that would make a survey such as this much more efficient.

\section{Data availability}

Data products described in Sect.~\ref{sec:dataproducts} will be made available through the ALMA Science Archive and/or the team webpage (\href{https://imager.oasu.u-bordeaux.fr}{https://imager.oasu.u-bordeaux.fr}). 

Table \ref{tab:spwinfo} is available with information for all sources in electronic form at the CDS via anonymous ftp to cdsarc.u-strasbg.fr (130.79.128.5) or via http://cdsweb.u-strasbg.fr/cgi-bin/qcat?J/A+A/.

\begin{acknowledgements}
We acknowledge the work of an anonymous referee that greatly improved the quality of this paper.  The authors express gratitude to the ALMA Observatory for enabling the successful execution of this Large Program and, in particular, staff at the Nordic ALMA Regional Centre node for their support during its planning and execution stages. S.M. and the team would like to thank St\'ephane Guilloteau for his invaluable help on the use of IMAGER for data reduction.  This paper makes use of the following ALMA data: ADS/JAO.ALMA\#2022.1.00316.L. ALMA is a partnership of ESO (representing its member states), NSF (USA) and NINS (Japan), together with NRC (Canada), NSTC and ASIAA (Taiwan), and KASI (Republic of Korea), in cooperation with the Republic of Chile. The Joint ALMA Observatory is operated by ESO, AUI/NRAO and NAOJ.  The National Radio Astronomy Observatory is a facility of the National Science Foundation operated under cooperative agreement by Associated Universities, Inc. The project leading to this publication has received support from ORP, that is funded by the European Union’s Horizon 2020 research and innovation programme under grant agreement No 101004719 [ORP].  Most of the computations presented in this paper were performed using the GRICAD infrastructure (\url{https://gricad.univ-grenoble-alpes.fr}), which is supported by Grenoble research communities.

We acknowledge financial support from the following: the European Research Council (ERC) under the European Union’s Horizon 2020 research and innovation programme (ERC Starting Grant ``Chemtrip,'' grant agreement No 949278; \emph{A.C., F.C.-S.M., A.A. and P.M.}); the National Research Foundation of Korea (NRF), funded by the Ministry of Science and ICT (MSIT) under grant numbers RS-2024-00416859 and RS-2026-25490557 (\emph{J.-E.L., J.-H.J., H.-S.Y., and C.-H.K.}), and by the Ministry of Education through the Basic Science Research Program under grant number RS-2025-25402622 (\emph{J.-H.J.}); the Thematic Action ``Physique et Chimie du Milieu Interstellaire'' (PCMI) of INSU Programme National ``Astro,'' with contributions from CNRS Physique \& CNRS Chimie, CEA, and CNES (\emph{S.M.}); the Ministry of Education of Taiwan (a grant from Center for Informatics and Computation in Astronomy, CICA,  and grant number 110J0353I9) and the National Science and Technology Council, Taiwan (Grant NSTC111-2112-M-007-014-MY3, NSTC113-2639-M-A49-002-ASP, and NSTC113-2112-M-007-027; \emph{D.H.}); the National Radio Astronomy Observatory and the North American ALMA Science Center (\emph{Z.T. and M.R.}); the NKFIH through NKKP grant ADVANCED 149943 and Excellence grant TKP2021-NKTA-64 (project no.149943 has been implemented with the support provided by the Ministry of Culture and Innovation of Hungary from the National Research, Development and Innovation Fund, financed under the NKKP ADVANCED funding scheme) as well as COST (European Cooperation in Science and Technology) through COST Action PLANETS CA22133 (A.K.); the Max Planck Society (\emph{S.S.}); the ESO and IAU Gruber Foundation Fellowship programs (\emph{P.N.}); the Ministry of Education, Culture, Sports, Science, and Technology of Japan (``Grant-in-Aid''; 20H05845, 20H05844) and RIKEN (pioneering project ``Evolution of Matter in the Universe''; \emph{(Y.-L.Y.)}; RIKEN through the Special Postdoctoral Researcher Program (Fellowships; \emph{J.F.A. and S.Z.}).
\end{acknowledgements}
\vspace{-1cm}
\bibliography{aa58751-25}{}
\bibliographystyle{aa}


\begin{appendix}

\onecolumn
\section{Observing logs} \label{sec:AAobslogs}
In this section, we include tabulated observing and imaging information.  Table \ref{tab:obspipelinelog} shows the observing logs for all SBs.  Start and end dates of observations for each SB are also represented graphically in Fig. \ref{fig:obstimeline}.  In very few cases (one tuning each for Serpens sources, Orion sources, and B335 IRS; and several tunings for NGC  1333-IRAS4A), an SB was observed with a configuration that differed from the ``nominal'' configuration used for the majority of the other SBs for a given source (shown with shaded vertical bands in Fig.~\ref{fig:obstimeline}); these corresponded to slightly longer maximum baselines of $\sim1.2-1.4$\ km, as indicated in Table \ref{tab:obspipelinelog}, whereas most observations for these sources were done with configurations with maximum baselines of $\sim700-800$\ m.  The corresponding effect of the smaller beam sizes for these specific tunings are apparent in Fig.~\ref{fig:continuumintensityandbeam}. NGC 1333-SVS13A and L1551-IRS5 were observed with configurations with maximum baselines of $\sim2.3-2.5$\ km.

Table \ref{tab:spwinfo} lists the imaging parameters for each target field, setting, and spectral window.  This table is intended as a guide to direct someone to the correct setting in case they want to look into a specific frequency range, and also to tabulate the synthesized beam and RMS noise for each dataset.  The representative values for each source that are given in Table \ref{tab:imaging2} are derived from the values in this table.

\setlength{\tabcolsep}{3pt} 
\begin{table}
\caption{Observing and pipeline log.}
\label{tab:obspipelinelog}
\centering
\tiny
\renewcommand{\arraystretch}{0.95}
\begin{tabular}{lllllllll}
\hline\hline
SB Name & MOUS & Calibrators & Obs. Start & Obs. End & N$_{\textrm{Ant}}$ & Max. BL & Pipeline Ver. & CASA Ver.\\
\hline
\multicolumn{9}{c}{\textbf{BHR71-IRS1}}\\
BHR71-IR\_a\_07\_TM1 & X2f7f & J1147-6753, J1427-4206 & 2023-01-08 & 2023-01-08 & 42 & 740.5 m & 2022.2.0.64 & 6.4.1.12 \\
BHR71-IR\_b\_07\_TM1 & X2f8f & J1147-6753, J1427-4206 & 2023-01-08 & 2023-01-08 & 42 & 740.5 m & 2022.2.0.64 & 6.4.1.12 \\
BHR71-IR\_c\_07\_TM1 & X2f9f & J1147-6753, J1206-6138, J1427-4206 & 2023-01-15 & 2023-01-15 & 46 & 783.5 m & 2022.2.0.64 & 6.4.1.12 \\
BHR71-IR\_d\_07\_TM1 & X2faf & J1147-6753, J1427-4206 & 2023-01-16 & 2023-01-16 & 46 & 783.5 m & 2022.2.0.64 & 6.4.1.12 \\
BHR71-IR\_e\_07\_TM1 & X2fbf & J1147-6753, J1427-4206 & 2023-01-16 & 2023-01-16 & 46 & 783.5 m & 2022.2.0.64 & 6.4.1.12 \\
BHR71-IR\_f\_07\_TM1 & X2fcf & J1147-6753, J1427-4206 & 2023-01-16 & 2023-01-16 & 46 & 783.5 m & 2022.2.0.64 & 6.4.1.12 \\
BHR71-IR\_g\_07\_TM1 & X2fdf & J1147-6753, J1427-4206 & 2023-01-17 & 2023-01-17 & 46 & 782.1 m & 2022.2.0.64 & 6.4.1.12 \\
BHR71-IR\_h\_07\_TM1 & X2fef & J1147-6753, J1427-4206 & 2023-01-22 & 2023-01-22 & 46 & 740.5 m & 2022.2.0.64 & 6.4.1.12 \\
BHR71-IR\_i\_07\_TM1 & X2fff & J1147-6753, J1427-4206 & 2023-01-22 & 2023-01-22 & 46 & 740.5 m & 2022.2.0.64 & 6.4.1.12 \\
\hline
\multicolumn{9}{c}{\textbf{B335 IRS}}\\
B335\_a\_07\_TM1 & X2f82\tablefootmark{a} & J1924-2914, J1938+0448 & 2024-05-09 & 2024-05-09 & 45 & 696.9 m & 2023.1.0.124 & 6.5.4.9 \\
B335\_b\_07\_TM1 & X2f92 & J1924+1540, J1924-2914 & 2023-04-12 & 2023-04-12 & 44 & 1.3 km & 2022.2.0.64 & 6.4.1.12 \\
B335\_c\_07\_TM1 & X2fa2 & J1924-2914, J1938+0448 & 2024-05-09 & 2024-05-09 & 45 & 696.9 m & 2023.1.0.124 & 6.5.4.9 \\
B335\_d\_07\_TM1 & X2fb2 & J1924-2914, J1938+0448 & 2024-05-08 & 2024-05-09 & 45 & 696.9 m & 2023.1.0.124 & 6.5.4.9 \\
B335\_e\_07\_TM1 & X2fc2 & J1924+1540, J1924-2914 & 2023-03-02 & 2023-03-02 & 41 & 783.5 m & 2022.2.0.64 & 6.4.1.12 \\
B335\_f\_07\_TM1 & X2fd2 & J1924+1540, J1924-2914, J1925+2106 & 2023-03-03 & 2023-03-03 & 41 & 783.5 m & 2022.2.0.64 & 6.4.1.12 \\
B335\_g\_07\_TM1 & X2fe2 & J1924+1540, J1924-2914 & 2023-03-03 & 2023-03-03 & 41 & 783.5 m & 2022.2.0.64 & 6.4.1.12 \\
B335\_h\_07\_TM1 & X2ff2 & J1427-4206, J1938+0448, J1951-0509 & 2024-01-02 & 2024-05-09 & 43 & 783.5 m & 2023.1.0.124 & 6.5.4.9 \\
B335\_i\_07\_TM1 & X3002 & J1924-2914, J1938+0448 & 2024-05-08 & 2024-05-08 & 46 & 696.9 m & 2023.1.0.124 & 6.5.4.9 \\
\hline
\multicolumn{9}{c}{\textbf{HOPS 108, HOPS 373-B, V883-Ori }}\\
HOPS-108\_a\_07\_TM1 & X2f9c & J0423-0120, J0529-0519 & 2024-01-08 & 2024-05-19 & 43 & 696.9 m & 2023.1.0.124 & 6.5.4.9 \\
HOPS-108\_b\_07\_TM1 & X2fac & J0423-0120, J0529-0519 & 2024-01-08 & 2024-01-08 & 43 & 696.9 m & 2023.1.0.124 & 6.5.4.9 \\
HOPS-108\_bx\_07\_TM1 & X73\tablefootmark{b} & J0423-0120, J0529-0519 & 2024-08-08 & 2024-08-13 & 46 & 1.3 km & 2023.1.0.124 & 6.5.4.9 \\
HOPS-108\_c\_07\_TM1 & X2fcc & J0541-0541, J0607-0834, J0854+2006 & 2023-01-13 & 2023-01-17 & 47 & 783.5 m & 2022.2.0.64 & 6.4.1.12 \\
HOPS-108\_d\_07\_TM1 & X2fdc & J0423-0120, J0541-0541, J0607-0834 & 2023-01-23 & 2023-03-04 & 47 & 783.5 m & 2022.2.0.64 & 6.4.1.12 \\
HOPS-373\_a\_07\_TM1 & X2f7c & J0423-0120, J0541-0541 & 2023-01-13 & 2024-01-02 & 47 & 783.5 m & 2023.1.0.124 & 6.5.4.9 \\
HOPS-373\_b\_07\_TM1 & X2f8c & J0423-0120, J0529-0519 & 2024-01-02 & 2024-05-25 & 46 & 783.5 m & 2023.1.0.124 & 6.5.4.9 \\
HOPS-373\_c\_07\_TM1 & X2fec & J0541-0541, J0607-0834, J0854+2006 & 2023-01-09 & 2023-01-12 & 43 & 740.5 m & 2022.2.0.64 & 6.4.1.12 \\
HOPS-373\_d\_07\_TM1 & X2ffc & J0541-0541, J0607-0834, J0854+2006 & 2023-01-12 & 2023-01-15 & 47 & 783.5 m & 2022.2.0.64 & 6.4.1.12 \\
V883\_Ori\_a\_07\_TM1 & X2fbc & J0541-0541, J0607-0834, J0854+2006 & 2023-03-05 & 2024-01-03 & 44 & 783.5 m & 2023.1.0.124 & 6.5.4.9 \\
\hline
\multicolumn{9}{c}{\textbf{L1551-IRS5}}\\
L1551\_IR\_a\_07\_TM1 & X300c & J0423-0120, J0431+1731, J0440+1437 & 2024-06-26 & 2024-06-27 & 42 & 2.5 km & 2023.1.0.124 & 6.5.4.9 \\
L1551\_IR\_b\_07\_TM1 & X3013 & J0423-0120, J0431+1731, J0440+1437 & 2023-12-07 & 2023-12-07 & 44 & 2.5 km & 2023.1.0.124 & 6.5.4.9 \\
L1551\_IR\_c\_07\_TM1 & X301a & J0423-0120, J0431+1731, J0440+1437 & 2024-06-27 & 2024-06-28 & 42 & 2.5 km & 2023.1.0.124 & 6.5.4.9 \\
L1551\_IR\_d\_07\_TM1 & X3021 & J0423-0120, J0431+1731, J0440+1437 & 2024-06-24 & 2024-06-27 & 43 & 2.3 km & 2023.1.0.124 & 6.5.4.9 \\
L1551\_IR\_e\_07\_TM1 & X3028 & J0423-0120, J0431+1731, J0440+1437 & 2024-06-29 & 2024-06-29 & 42 & 2.5 km & 2023.1.0.124 & 6.5.4.9 \\
L1551\_IR\_f\_07\_TM1 & X302f & J0423-0120, J0431+1731, J0440+1437 & 2023-12-10 & 2023-12-10 & 46 & 2.5 km & 2023.1.0.124 & 6.5.4.9 \\
L1551\_IR\_g\_07\_TM1 & X3036 & J0423-0120, J0431+1731, J0440+1437 & 2024-06-28 & 2024-06-28 & 44 & 2.5 km & 2023.1.0.124 & 6.5.4.9 \\
L1551\_IR\_h\_07\_TM1 & X303d & J0423-0120, J0431+1731, J0440+1437 & 2024-06-29 & 2024-07-14 & 42 & 2.5 km & 2023.1.0.124 & 6.5.4.9 \\
L1551\_IR\_i\_07\_TM1 & X3044 & J0423-0120, J0431+1731, J0440+1437 & 2023-12-09 & 2023-12-10 & 47 & 2.5 km & 2023.1.0.124 & 6.5.4.9 \\
\hline
\multicolumn{9}{c}{\textbf{NG1333-IRAS4A }}\\
NGC1333\_a\_07\_TM1 & X2f79 & J0336+3218, J0357+2319, J0423-0120 & 2023-01-09 & 2023-01-09 & 41 & 740.5 m & 2022.2.0.64 & 6.4.1.12 \\
NGC1333\_b\_07\_TM1 & X2f89 & J0336+3218, J0423-0120 & 2023-12-24 & 2023-12-24 & 43 & 1.2 km & 2023.1.0.124 & 6.5.4.9 \\
NGC1333\_c\_07\_TM1 & X2f99 & J0336+3218, J0423-0120 & 2023-12-22 & 2023-12-22 & 46 & 1.4 km & 2023.1.0.124 & 6.5.4.9 \\
NGC1333\_d\_07\_TM1 & X2fa9 & J0237+2848, J0336+3218, J0423-0120 & 2024-07-31 & 2024-07-31 & 50 & 1.4 km & 2023.1.0.124 & 6.5.4.9 \\
NGC1333\_e\_07\_TM1 & X2fb9 & J0336+3218, J0357+2319, J0423-0120 & 2024-01-03 & 2024-08-02 & 45 & 783.5 m & 2023.1.0.124 & 6.5.4.9 \\
NGC1333\_f\_07\_TM1 & X2fc9 & J0325+2224, J0336+3218, J0423-0120 & 2024-08-01 & 2024-08-01 & 48 & 1.4 km & 2023.1.0.124 & 6.5.4.9 \\
NGC1333\_g\_07\_TM1 & X2fd9 & J0336+3218, J0357+2319, J0423-0120 & 2024-08-02 & 2024-08-03 & 50 & 1.4 km & 2023.1.0.124 & 6.5.4.9 \\
NGC1333\_h\_07\_TM1 & X2fe9 & J0336+3218, J0357+2319, J0423-0120 & 2023-12-23 & 2023-12-23 & 44 & 1.2 km & 2023.1.0.124 & 6.5.4.9 \\
NGC1333\_i\_07\_TM1 & X2ff9 & J0336+3218, J0357+2319, J0423-0120 & 2023-12-23 & 2023-12-23 & 44 & 1.2 km & 2023.1.0.124 & 6.5.4.9 \\
\hline
\multicolumn{9}{c}{\textbf{NGC1333-SVS13a }}\\
SVS13-A\_a\_07\_TM1 & X3009 & J0328+3139, J0336+3218, J0423-0120 & 2023-12-05 & 2023-12-05 & 43 & 2.5 km & 2023.1.0.124 & 6.5.4.9 \\
SVS13-A\_b\_07\_TM1 & X3010 & J0325+2224, J0328+3139, J0336+3218, J0423-0120 & 2024-06-27 & 2024-06-27 & 45 & 2.5 km & 2023.1.0.124 & 6.5.4.9 \\
SVS13-A\_c\_07\_TM1 & X3017 & J0328+3139, J0336+3218, J0423-0120 & 2023-12-10 & 2023-12-10 & 46 & 2.5 km & 2023.1.0.124 & 6.5.4.9 \\
SVS13-A\_d\_07\_TM1 & X301e & J0328+3139, J0336+3218, J0423-0120 & 2023-12-06 & 2023-12-09 & 46 & 2.5 km & 2023.1.0.124 & 6.5.4.9 \\
SVS13-A\_e\_07\_TM1 & X3025 & J0325+2224, J0328+3139, J0336+3218, J0423-0120 & 2024-06-28 & 2024-06-28 & 45 & 2.5 km & 2023.1.0.124 & 6.5.4.9 \\
SVS13-A\_f\_07\_TM1 & X302c & J0325+2224, J0328+3139, J0336+3218, J0423-0120 & 2024-06-26 & 2024-06-26 & 43 & 2.4 km & 2023.1.0.124 & 6.5.4.9 \\
SVS13-A\_g\_07\_TM1 & X3033 & J0328+3139, J0336+3218, J0423-0120 & 2024-06-26 & 2024-06-26 & 42 & 2.5 km & 2023.1.0.124 & 6.5.4.9 \\
SVS13-A\_h\_07\_TM1 & X303a & J0336+3218, J0357+2319, J0403+2600, J0423-0120 & 2023-12-06 & 2023-12-06 & 46 & 2.5 km & 2023.1.0.124 & 6.5.4.9 \\
SVS13-A\_i\_07\_TM1 & X3041 & J0328+3139, J0336+3218, J0423-0120 & 2023-12-07 & 2023-12-07 & 44 & 2.5 km & 2023.1.0.124 & 6.5.4.9 \\
\hline
\multicolumn{9}{c}{\textbf{Serpens emb 11 W, Serpens South-MM18a, Serpens SMM1a }}\\
Ser\_emb\_a\_07\_TM1 & X2ff5 & J1851+0035, J1924-2914 & 2024-05-13 & 2024-05-13 & 46 & 783.5 m & 2023.1.0.124 & 6.5.4.9 \\
Ser\_emb\_b\_07\_TM1 & X3005 & J1851+0035, J1924-2914 & 2024-05-11 & 2024-05-12 & 47 & 783.5 m & 2023.1.0.124 & 6.5.4.9 \\
SerpS-MM\_a\_07\_TM1 & X2f85 & J1851+0035, J1924-2914 & 2024-05-14 & 2024-05-21 & 46 & 783.5 m & 2023.1.0.124 & 6.5.4.9 \\
SerpS-MM\_b\_07\_TM1 & X2fa5 & J1851+0035, J1924-2914 & 2024-05-25 & 2024-05-27 & 41 & 740.5 m & 2023.1.0.124 & 6.5.4.9 \\
SerpS-MM\_c\_07\_TM1 & X2fb5 & J1851+0035, J1924-2914 & 2024-05-13 & 2024-05-21 & 46 & 783.5 m & 2023.1.0.124 & 6.5.4.9 \\
SerpS-MM\_d\_07\_TM1 & X2fc5 & J1851+0035, J1924-2914 & 2024-05-29 & 2024-07-29 & 44 & 1.3 km & 2023.1.0.124 & 6.5.4.9 \\
SerpS-MM\_e\_07\_TM1 & X2fd5 & J1851+0035, J1924-2914 & 2024-05-12 & 2024-05-12 & 46 & 783.5 m & 2023.1.0.124 & 6.5.4.9 \\
Serp\_SMM\_a\_07\_TM1 & X2f95 & J1851+0035, J1924-2914 & 2024-05-24 & 2024-05-24 & 43 & 740.5 m & 2023.1.0.124 & 6.5.4.9 \\
Serp\_SMM\_b\_07\_TM1 & X2fe5 & J1851+0035, J1924-2914 & 2024-05-11 & 2024-05-11 & 46 & 783.5 m & 2023.1.0.124 & 6.5.4.9 \\
\hline

\end{tabular}
\tablefoot{For each target and scheduling block (SB) name, columns show the MOUS ID, calibrators, observation start and end dates, number of antennas available (N$_\textrm{Ant}$), maximum baseline length, ALMA Pipeline version and CASA version used for data calibration.
\tablefoottext{a}{Full MOUS UID prefix is uid://A001/X2d20/ for all SBs in this project, except where indicated.}
\tablefoottext{b}{Full MOUS UID is uid://A001/X3734/ for this EB, which is different from the others since it was reobserved at a later date.}
\tablefoottext{c}{Sources listed are phase, bandpass, and amplitude calibrators.}
}
\end{table}

\begin{table}[h!]
\caption{Imaging parameters by SPW.}
\label{tab:spwinfo}
\centering
\tiny
\begin{tabular}{cccccc}
\hline\hline
Tuning & SPW & Min.~freq.& Max.~freq.& HPBW (PA) & $\sigma_{\textrm{RMS,\ cube}}$ \\
 & & $\mathrm{(GHz)}$ & $\mathrm{(GHz)}$ & & $\mathrm{(mJy\,bm^{-1}\,ch^{-1})}$ \\
\hline
\multicolumn{6}{c}{\textbf{BHR71-IRS1}}\\
1 & 25 & 279.0444 & 279.9794 & 0.51$''$$\times$0.38$''$ ( 22$^\circ$) & 2.17 \\
 & 27 & 279.9519 & 280.8869 & 0.51$''$$\times$0.37$''$ ( 23$^\circ$) & 2.14 \\
 & 29 & 280.8594 & 281.7944 & 0.50$''$$\times$0.37$''$ ( 24$^\circ$) & 2.08 \\
 & 31 & 281.7669 & 282.7019 & 0.50$''$$\times$0.37$''$ ( 24$^\circ$) & 2.12 \\
\arrayrulecolor{lightgray}\hline 
2 & 25 & 282.6744 & 283.6094 & 0.52$''$$\times$0.35$''$ ( 45$^\circ$) & 2.19 \\
 & 27 & 283.5819 & 284.5169 & 0.51$''$$\times$0.35$''$ ( 45$^\circ$) & 2.17 \\
 & 29 & 284.4894 & 285.4244 & 0.51$''$$\times$0.35$''$ ( 45$^\circ$) & 2.13 \\
 & 31 & 285.3969 & 286.3319 & 0.50$''$$\times$0.35$''$ ( 45$^\circ$) & 2.59 \\
\arrayrulecolor{lightgray}\hline 
3 & 25 & 286.3044 & 287.2394 & 0.44$''$$\times$0.35$''$ (155$^\circ$) & 1.88 \\
 & 27 & 287.2119 & 288.1469 & 0.44$''$$\times$0.35$''$ (156$^\circ$) & 1.89 \\
 & 29 & 288.1194 & 289.0544 & 0.44$''$$\times$0.34$''$ (156$^\circ$) & 2.08 \\
 & 31 & 289.0269 & 289.9619 & 0.44$''$$\times$0.34$''$ (155$^\circ$) & 1.91 \\
\arrayrulecolor{lightgray}\hline 
4 & 25 & 289.9341 & 290.8692 & 0.44$''$$\times$0.35$''$ (172$^\circ$) & 1.87 \\
 & 27 & 290.8416 & 291.7766 & 0.43$''$$\times$0.35$''$ (173$^\circ$) & 1.86 \\
 & 29 & 291.7491 & 292.6842 & 0.43$''$$\times$0.35$''$ (174$^\circ$) & 1.95 \\
 & 31 & 292.6566 & 293.5917 & 0.44$''$$\times$0.35$''$ (175$^\circ$) & 2.06 \\
\arrayrulecolor{lightgray}\hline 
5 & 25 & 293.5641 & 294.4992 & 0.42$''$$\times$0.33$''$ ( 17$^\circ$) & 2.13 \\
 & 27 & 294.4716 & 295.4066 & 0.42$''$$\times$0.33$''$ ( 17$^\circ$) & 2.09 \\
 & 29 & 295.3791 & 296.3142 & 0.42$''$$\times$0.33$''$ ( 16$^\circ$) & 2.15 \\
 & 31 & 296.2866 & 297.2217 & 0.42$''$$\times$0.33$''$ ( 16$^\circ$) & 2.14 \\
\arrayrulecolor{lightgray}\hline 
6 & 25 & 297.1941 & 298.1292 & 0.46$''$$\times$0.34$''$ ( 44$^\circ$) & 2.06 \\
 & 27 & 298.1016 & 299.0366 & 0.46$''$$\times$0.34$''$ ( 44$^\circ$) & 2.21 \\
 & 29 & 299.0091 & 299.9442 & 0.45$''$$\times$0.34$''$ ( 44$^\circ$) & 2.24 \\
 & 31 & 299.9166 & 300.8517 & 0.45$''$$\times$0.34$''$ ( 44$^\circ$) & 2.30 \\
\arrayrulecolor{lightgray}\hline 
7 & 25 & 300.8241 & 301.7592 & 0.46$''$$\times$0.33$''$ ( 36$^\circ$) & 2.62 \\
 & 27 & 301.7316 & 302.6666 & 0.46$''$$\times$0.33$''$ ( 37$^\circ$) & 2.88 \\
 & 29 & 302.6391 & 303.5742 & 0.46$''$$\times$0.33$''$ ( 36$^\circ$) & 2.74 \\
 & 31 & 303.5466 & 304.4817 & 0.46$''$$\times$0.33$''$ ( 36$^\circ$) & 2.51 \\
\arrayrulecolor{lightgray}\hline 
8 & 25 & 304.4542 & 305.3892 & 0.45$''$$\times$0.33$''$ (  3$^\circ$) & 2.00 \\
 & 27 & 305.3616 & 306.2966 & 0.45$''$$\times$0.33$''$ (  4$^\circ$) & 2.08 \\
 & 29 & 306.2691 & 307.2042 & 0.46$''$$\times$0.32$''$ (  5$^\circ$) & 2.20 \\
 & 31 & 307.1766 & 308.1117 & 0.45$''$$\times$0.33$''$ (  4$^\circ$) & 2.16 \\
\arrayrulecolor{lightgray}\hline 
9 & 25 & 308.0842 & 309.0192 & 0.45$''$$\times$0.32$''$ ( 28$^\circ$) & 2.03 \\
 & 27 & 308.9916 & 309.9266 & 0.43$''$$\times$0.30$''$ ( 28$^\circ$) & 2.12 \\
 & 29 & 309.8991 & 310.8342 & 0.45$''$$\times$0.32$''$ ( 28$^\circ$) & 2.24 \\
 & 31 & 310.8066 & 311.7417 & 0.45$''$$\times$0.30$''$ ( 29$^\circ$) & 2.22 \\
\hline
\end{tabular}
\tablefoot{Partial table here is shown for the case of BHR71-IRS1.  The full table is available at the CDS, with respective values for each target in the sample. $\sigma_{\textrm{RMS,\ cube}}$ is the RMS noise for a channel width of 483.281 kHz (0.47-0.52 km s$^{-1}$).}
\end{table}

\twocolumn
\section{CASA imaging procedure} \label{sec:AAcasa}

\begin{figure}[!htb]
\begin{minipage}{0.48\textwidth}
\includegraphics[width=\textwidth]{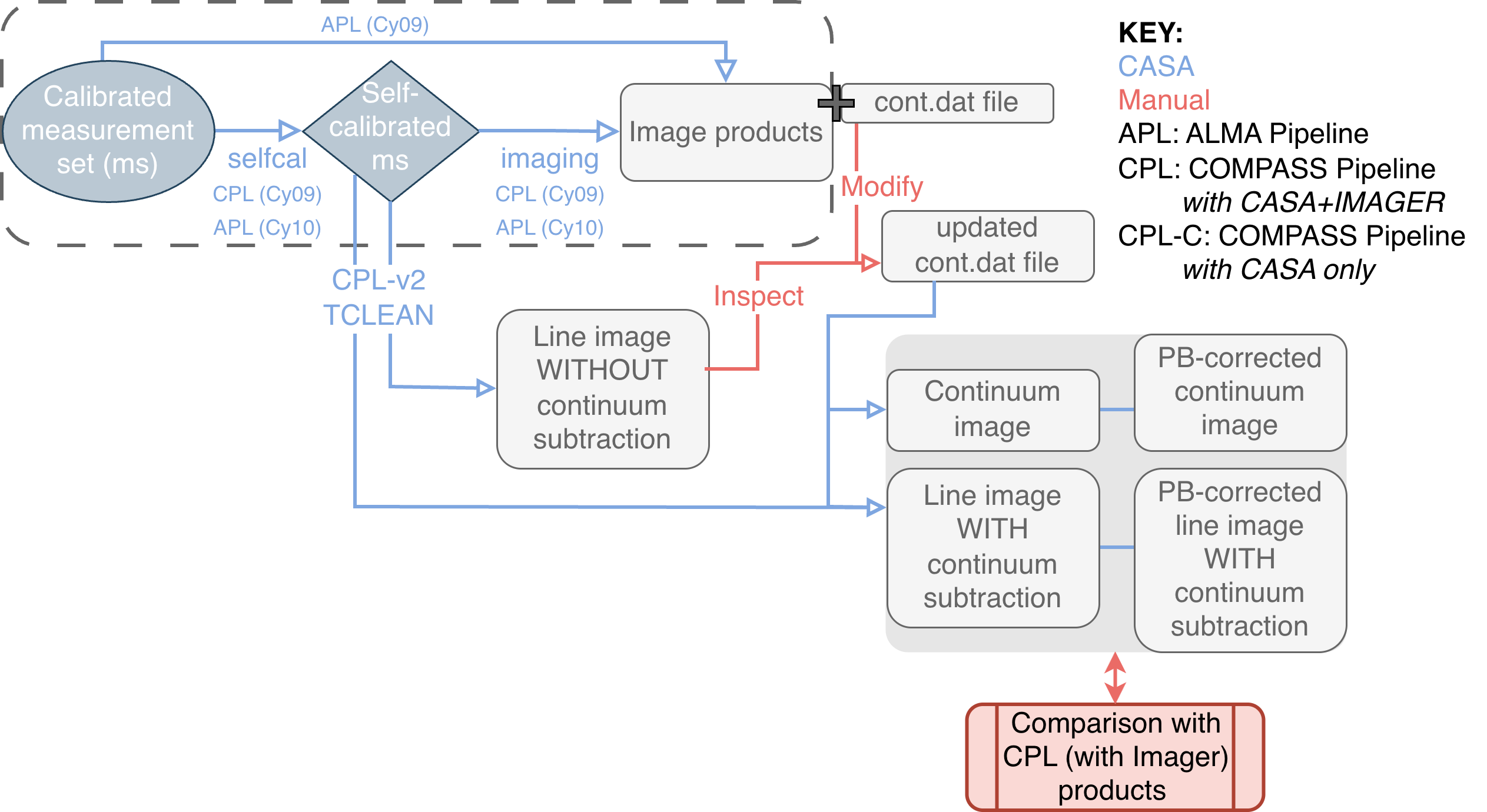}
\caption{CASA data flow, including a second variation of the COMPASS pipeline (referred to as CPL-C) that utilizes exclusively CASA and not IMAGER.  It provides a few modifications with respect to the ALMA Pipeline. The dashed-outlined region indicates steps that are part of the ALMA Pipeline, and identical to a region in the primary data flow depicted in Fig. \ref{fig:imagerflow} (see Sect.~\ref{sec:imaging}).
 }\label{fig:casaflow}
 \end{minipage}
\end{figure}

While the COMPASS Pipeline blends CASA and IMAGER tasks (see Fig.~\ref{fig:imagerflow}), followed by post-processing, an alternative version can be done completely with CASA.  We implemented this version in order to validate the COMPASS Pipeline and provide a benchmark in terms of computational time, which we refer to as COMPASS Pipeline-CASA (or CPL-C).  Here, we provide a brief explanation of the benchmark pipeline that might be beneficial for the reader who is more familiar with CASA. The complete CASA version of the pipeline was run for all settings of BHR71-IRS1, and it is described visually in Fig.~\ref{fig:casaflow}.  

Imaging in CASA was done following the standard ALMA Pipeline \citep{Hunter2023}, with a few slight modifications.  For this, we used the ALMA Imaging Pipeline Reprocessing Tool.\footnote{\url{https://casaguides.nrao.edu/index.php/ALMA_Imaging_Pipeline_Reprocessing_Tool}}  As for the ALMA Pipeline, masking was done automatically using the auto-multithresh algorithm; we chose to use Brigg's weighting with robust=0.5.  One modification we made was to image without continuum subtraction in order to carefully validate, by eye, the channels that were identified in the ALMA Pipeline as line-free (saved in the file ``cont.dat'').  We used the ``cont.dat'' file as a starting point, and then made manual modifications to either add or remove certain spectral ranges, with the aim to include the largest possible bandwidth of likely line-free channels that could be used for continuum subtraction. 

Another modification we made was with respect to the \texttt{TCLEAN} parameter ``threshold.'' The ALMA Pipeline uses a dynamic range heuristic to determine the imaging threshold (effectively, how deep to clean).  Our script included an additional \texttt{TCLEAN} command that cleaned each SPW to a set threshold of 3.8 mJy beam$^{-1}$, or $\sim2\sigma$, where $\sigma$\ was the theoretical RMS noise level.  This imparted consistency across the complete frequency range of the spectra, for later comparison. 

\section{IMAGER and CASA}  \label{sec:imagervscasa}
While CASA is the most commonly used software package for ALMA data processing, the CPL additionally makes use of IMAGER for critical components of the process (see Fig. \ref{fig:imagerflow}).  It is important to emphasize that we do not intend to suggest that IMAGER (as utilized in the CPL) is the only method to produce high quality image cubes, from which spectra such as those we require for our studies are extracted.  Instead, we aim to relay some of the considerations that led us to choose IMAGER for the imaging steps in our particular use case, so that the reader can make their own informed choice.    

A key distinction between the CPL and CPL-C pipelines is computational efficiency. For the BHR71-IRS1 observations, IMAGER required 40–65 minutes to self-calibrate and image both spectral line and continuum data for one setting (four SPWs) on a single computing node with 32 cores and 192 GB RAM\footnote{IMAGER supports shared memory parallelism, enabling simultaneous processing of multiple spectral channels on a single node with multiple cores.}. In contrast, the equivalent CASA processing (in CPL-C) required up to several days on a comparable node.  Ultimately, the time advantage of using IMAGER in the CPL was on the order of hour(s) rather than day(s) for each setting.  Because we needed to test pipelines with several versions of image generation, investigate the impact of self-calibration on image outcomes (Sect.~\ref{sec:selfcal}), and evaluate continuum identification methods (see Sect.~\ref{sec:continuumid}) across all the data, efficient processing was essential.

The CPL is structured as a set of scripts tailored to each source and frequency setting, enabling independent processing. This design supports distributed execution across multiple cluster nodes. The utility of an optimized computing approach was underscored later in the project when an updated version of IMAGER with an improved continuum identification algorithm became available, and we wanted to test it for completeness. We reprocessed all sources and spectral windows within a few days using the cluster, demonstrating the pipeline's scalability.  Because we needed to test pipelines with several different versions of image generation and continuum identification methods (see Sec. \ref{sec:continuumid}) across all the data, efficient processing was essential.

\begin{figure*}[t!] 
    \centering
    \begin{minipage}[t]{0.48\textwidth}
        \centering
        \includegraphics[width=\textwidth]{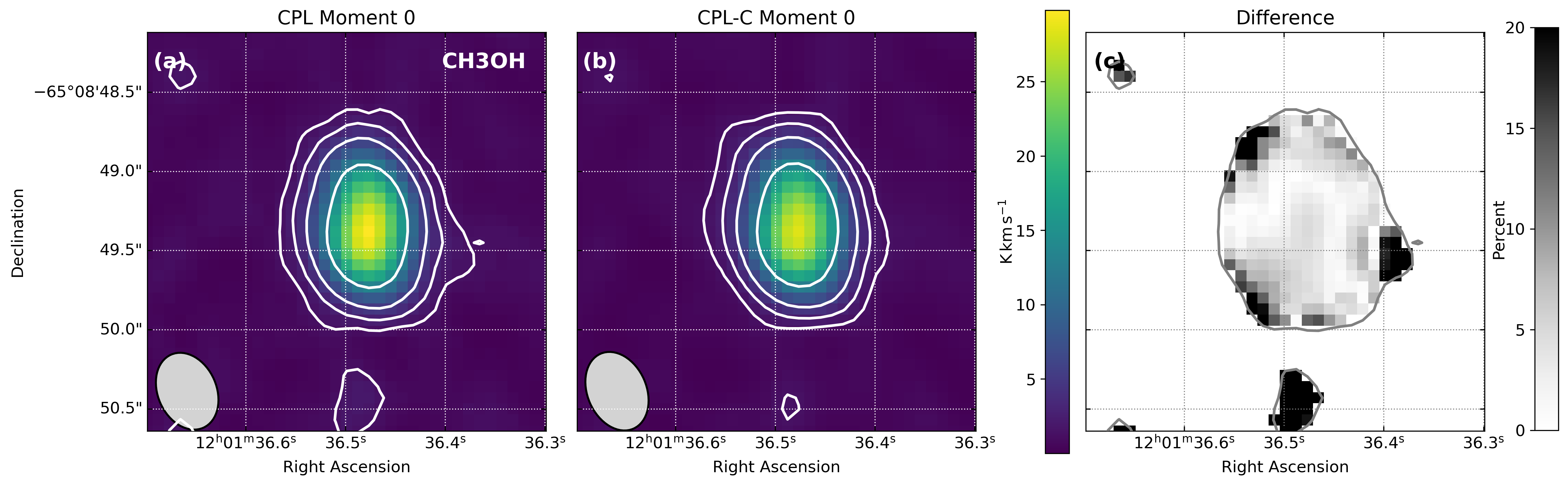}
        \includegraphics[width=\textwidth]{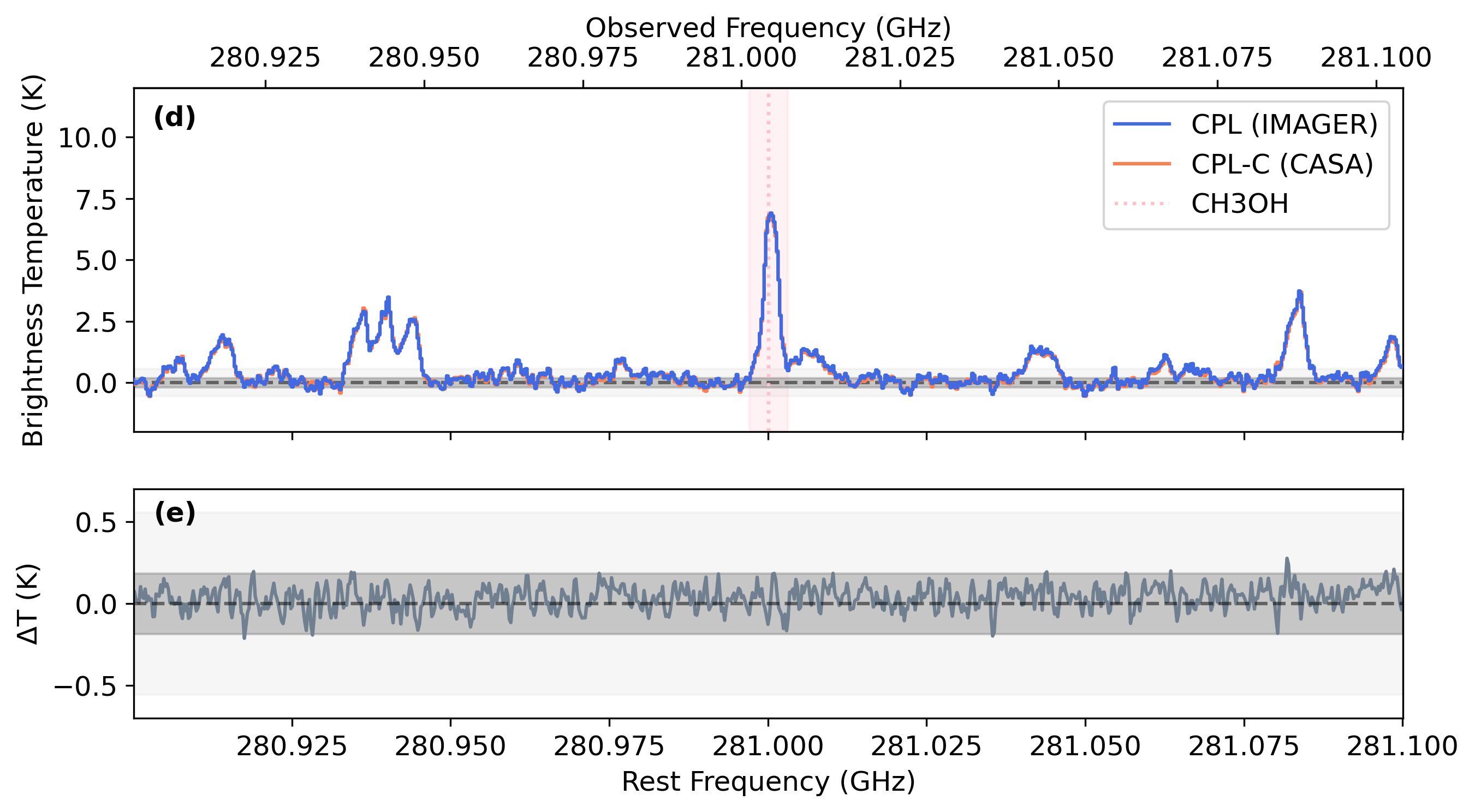} 
        \caption{Comparison between the zeroth-order moment maps of the CH$_3$OH $11(2)-10(3)$  transition at 281.0001 GHz produced with (a) CPL and (b) CPL-C in BHR71-IRS1. The white contours correspond to 5$\sigma$, 10$\sigma$, 20$\sigma$, 50$\sigma$, where $\sigma$ is the RMS noise level of the moment map in panel (a). The gray ellipses show the synthesized beams.  The percent difference in brightness temperature between the two images for each pixel is shown in (c); there, the gray contour is identical to the outer contour from (a).  (d) Comparison between spectra produced with the CPL (blue line) and CPL-C (orange line). Both spectra are extracted from the BHR71-IRS1 phase center (see Table \ref{tab:imaging2}). The vertical pink region is included for reference to indicate the frequency range over which moment 0 maps (above) were integrated. (e) The difference in brightness temperatures between the two spectra, with a difference below $\sim1\sigma_{\textrm{RMS,\ cube}}$ at nearly every frequency channel. The dark and light gray areas in (d) and (e) represent the $1\sigma_{\textrm{RMS,\ cube}}$ and $3\sigma_{\textrm{RMS,\ cube}}$ noise levels, respectively (see Table \ref{tab:spwinfo} for noise level of the respective tuning and SPW).}
        \label{fig:bhr71-CH3OH-comparison}
    \end{minipage}
    \hfill 
    \begin{minipage}[t]{0.48\textwidth}
        \centering
        \includegraphics[width=\textwidth]{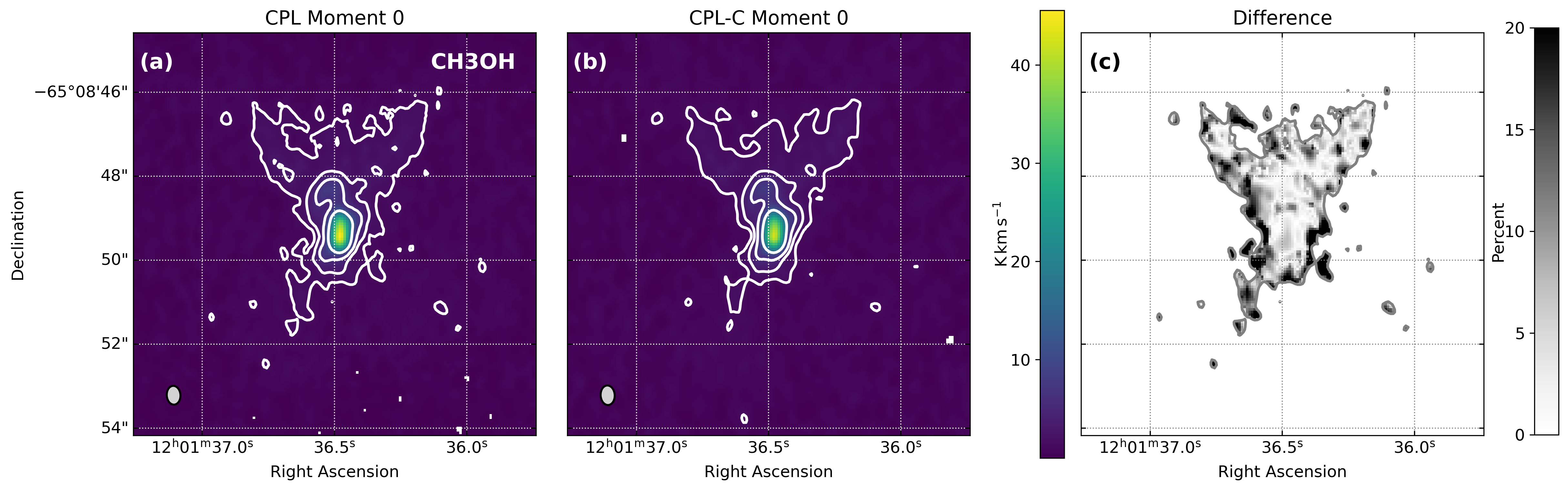}
        \includegraphics[width=\textwidth]{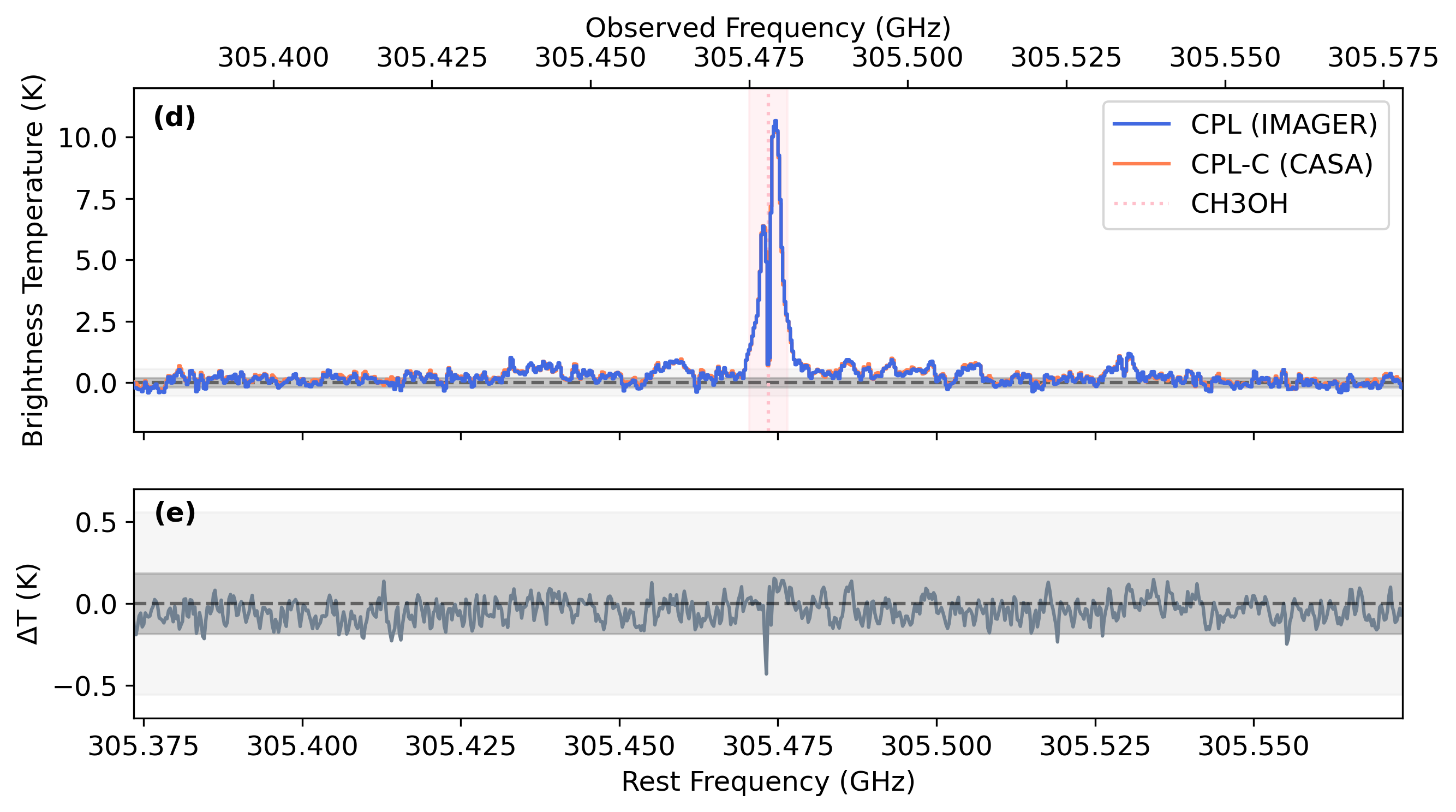} 
       \caption{Same as Fig.~\ref{fig:bhr71-CH3OH-comparison}, but for the CH$_3$OH $3(1,2)-3(0,3)$\ transition at 305.473491 GHz, in order to show the effect in the case of a spectral line with a more spatially extended morphology.}
        \label{fig:bhr71-13CH3OH-comparison}
    \end{minipage}
\end{figure*}

In order to validate the methods, we inspected comparable images and spectra from the CPL and CPL-C.  For example, in Figs.~\ref{fig:bhr71-CH3OH-comparison} and \ref{fig:bhr71-13CH3OH-comparison}, we compare the zeroth-order moment maps and spectra of two methanol transitions with different spatial extents, produced from the CPL and CPL-C.  Despite being obtained through independent imaging pipelines, the map morphologies (panels a-c) and spectra (panels d-e) are scientifically comparable. The spectra are vastly equivalent within less than the RMS noise of the respective data cubes.

\section{Alternative methods for continuum identification and subtraction } \label{sec:continuumalternatives}

In this section we outline the other algorithms and methods that were tested for continuum identification and subtraction, in the context of the CPL method (using IMAGER, and post-processing) and the CPL-C method (using CASA, and continuum subtraction in the $uv$\ plane).  First, several statistical methods exist, in addition to the method from \citet{jorgensen2016} that we ultimately adopted as CPL (see Sect.~\ref{sec:continuumid}). We tested two methods using IMAGER and continuum subtraction in the image plane, specifically using the algorithms called ``sigma-clipping method'' (SCM) and ``corrected sigma-clipping method'' (C-SCM).  These are conceptually similar to the Python-based STATCONT method thoroughly described by \citet{Sanchez2018}, which we also tested and found comparable to the previously mentioned IMAGER methods.  

On the other hand, these are conceptually different from those methods that depend on identifying certain channels that have no significant contribution from spectral lines.  With such identification, the subtraction could be done in the $uv$\ or image plane. In the CPL-C method that we tested (see Appendix \ref{sec:AAcasa}), line-free channels are identified in the images, and applied in the $uv$\ plane; IMAGER also has an option to perform continuum subtraction in the $uv$\ plane, which we employed for generating continuum images. 

Figure \ref{fig:contsub} shows comparisons of spectra obtained from several of the different continuum subtraction and imaging methods mentioned here; the post-processing (CPL) data is generally more consistent with the CASA (CPL-C) version than with either of the versions that used IMAGER for the continuum subtraction. Overall, CPL or CPL-C datasets should provide scientifically equivalent results for most science cases.

\begin{figure*}[t!] 
    \centering
    \begin{minipage}[t]{0.47\textwidth}
        \centering
        \includegraphics[width=\textwidth]{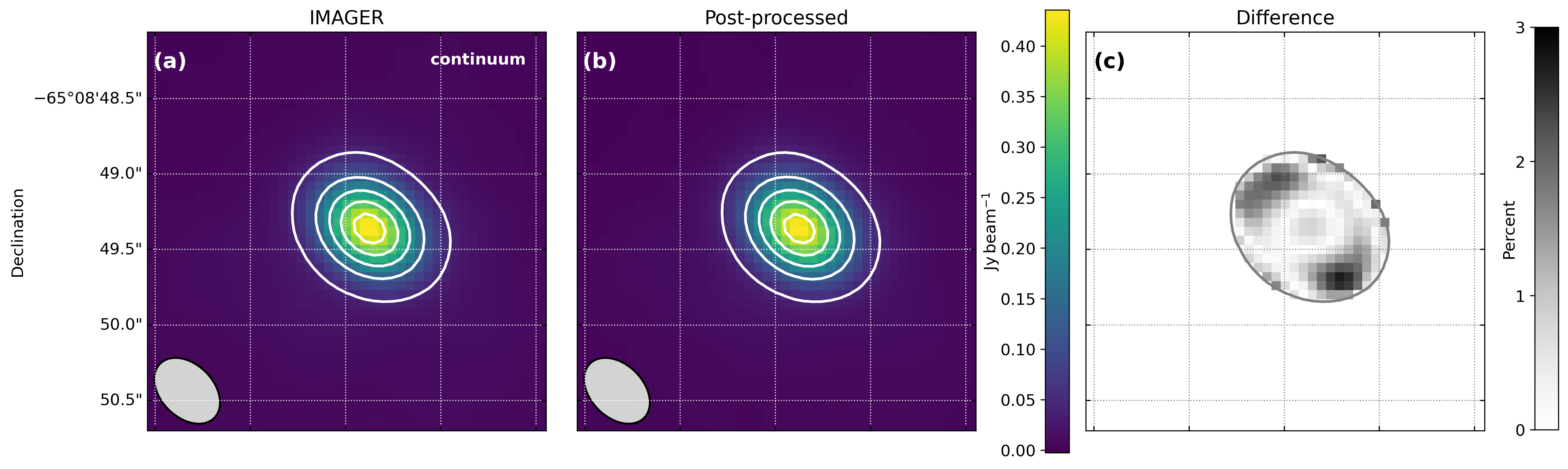}
        \includegraphics[width=\textwidth]{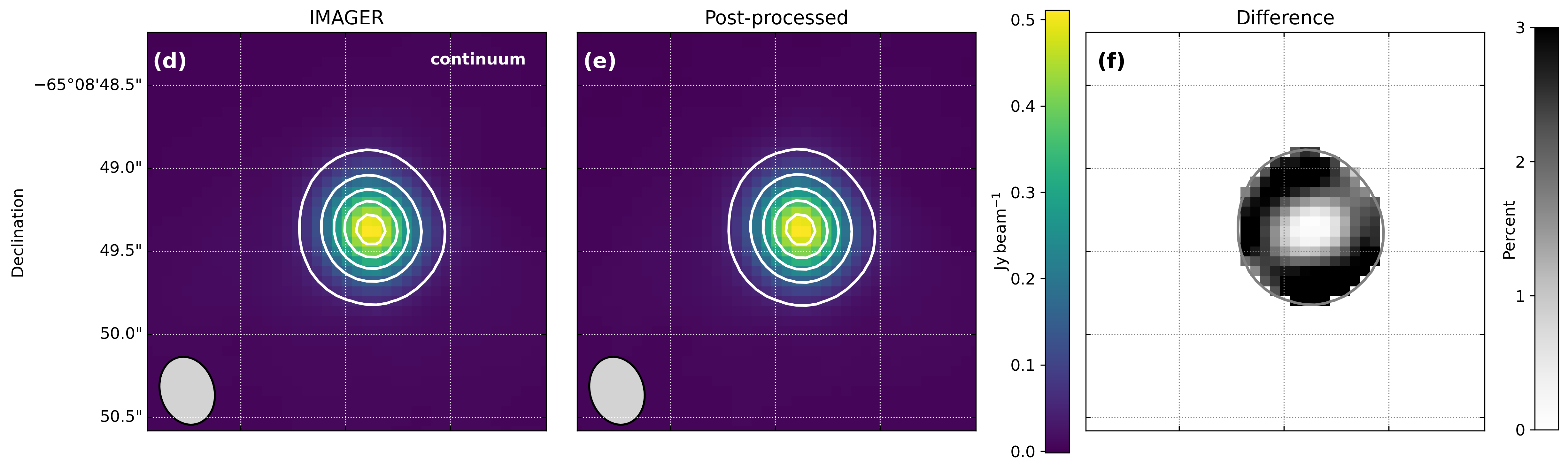}
        \includegraphics[width=\textwidth]{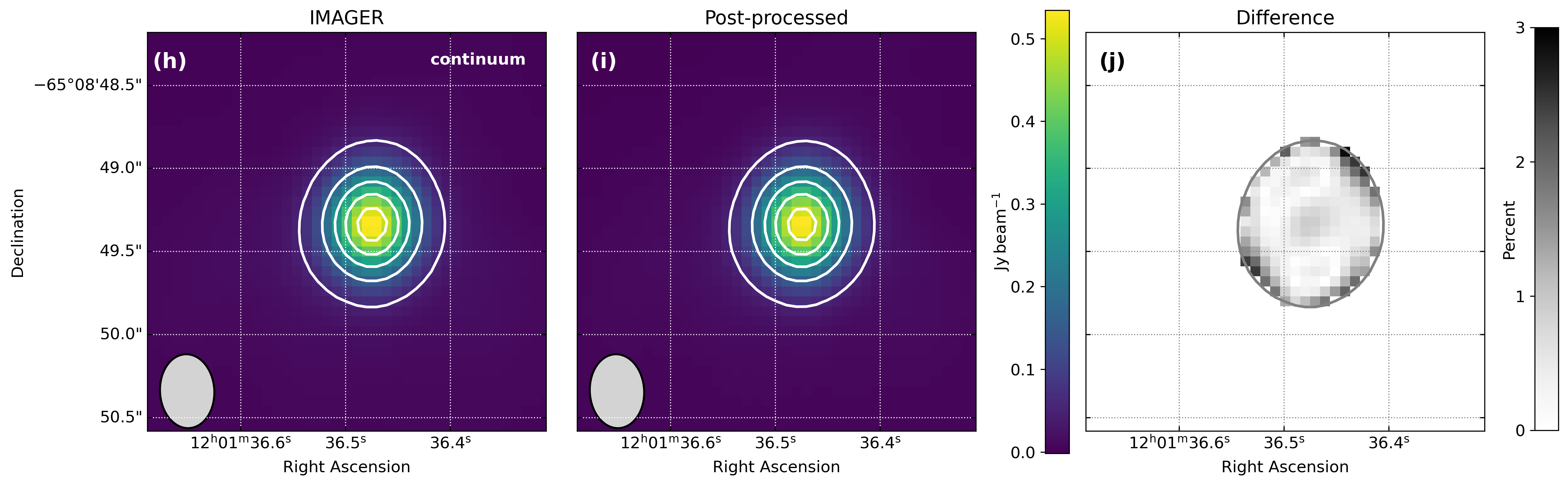}
        \caption{Comparisons between continuum images of BHR71-IRS1 generated using two different methods presented in this paper, with upper, middle, and lower rows representing spectral tunings 2, 5, and 8, respectively.  The right column shows the percent difference in all pixels with an intensity greater than 10\% of the peak value; differences are less than 4\%.} \label{fig:cont_compare}
    \end{minipage}
    \hfill 
    \begin{minipage}[t]{0.51\textwidth}
        \centering
        \includegraphics[width=\textwidth]{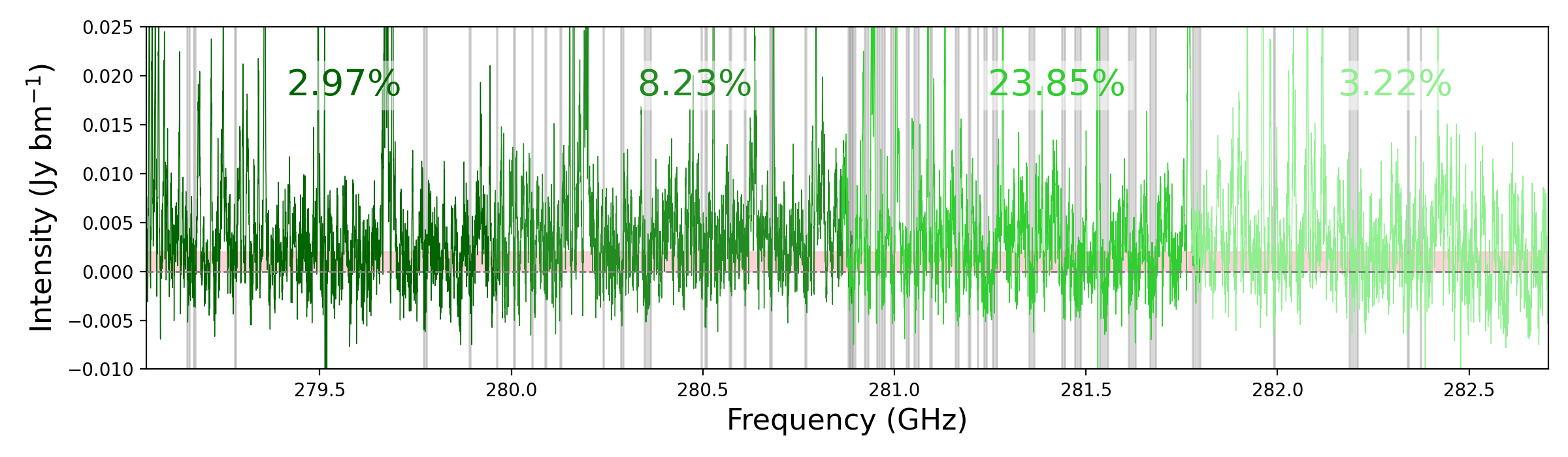}
        \includegraphics[width=\textwidth]{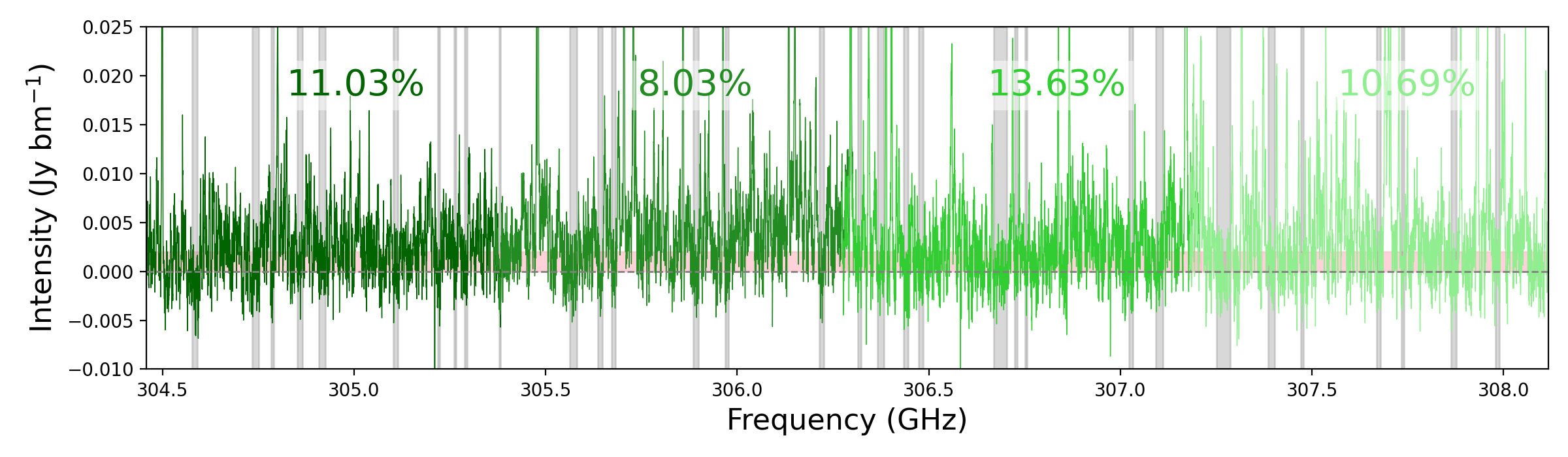}
        \caption{Spectra for BHR71-IRS1, tunings 1 and 8, with continuum channels identified by the CPL-C (see Fig.~\ref{fig:casaflow}) shown with vertical gray shading.  These representative spectra were extracted from the images generated by the CPL-C, following the method in Appendix \ref{sec:AAcasa}. The percentage of channels identified as continuum in each SPW labeled; tones of green represent the four SPWs for each setting.  In many cases across the full COMPASS dataset, the identified continuum range is too limited ($\lesssim10\%$) to make a precise continuum subtraction without using an advanced statistical method.  The strongest emission features shown here extend beyond the y-axis range selected for this plot.  The horizontal light pink band indicates the $1\sigma$\ level of 2 mJy bm$^{-1}$.} \label{fig:contid}
    \end{minipage}
\end{figure*}

While this paper focuses on science cases related to the spectral lines, for which the observations had been set up accordingly, we acknowledge that the method we adopted for generating continuum images could result in a relative over-estimation of the continuum fluxes. We show examples of continuum images that we obtain for several spectral tunings in Fig.~\ref{fig:cont_compare}, panels (a), (d), and (h), and compare these with the respective images produced as a byproduct of the CPL method in panels (b), (e), and (i) (see Sec.~\ref{sec:continuumid}).  The possible over-estimation is less than a few percent, shown in panels (c), (f), and (j).

As part of the ALMA pipeline and CPL-C, frequency ranges that are identified as free of spectral line emission are listed in a file called ``cont.dat'' for each SPW.  The continuum subtraction that is done in the $uv$\ plane takes into account these frequency ranges.  We highlight in Fig. \ref{fig:contid} that a very small percentage of the spectral range is identified as line-free, and the accuracies of the identifications were hard to validate, even with a careful ``by eye'' assessment. This led us to choose a statistical method; readers can refer to Sect.~\ref{sec:continuumid} for further discussion of the method utilized for our study.

\newpage
\section{Self-calibration improvements of the continuum images 
  \label{sec:impr-self-calibr}}

\begin{figure*}[t!] 
    \centering
    \begin{minipage}[t]{0.64\textwidth}
        \centering
        \includegraphics[width=\textwidth]{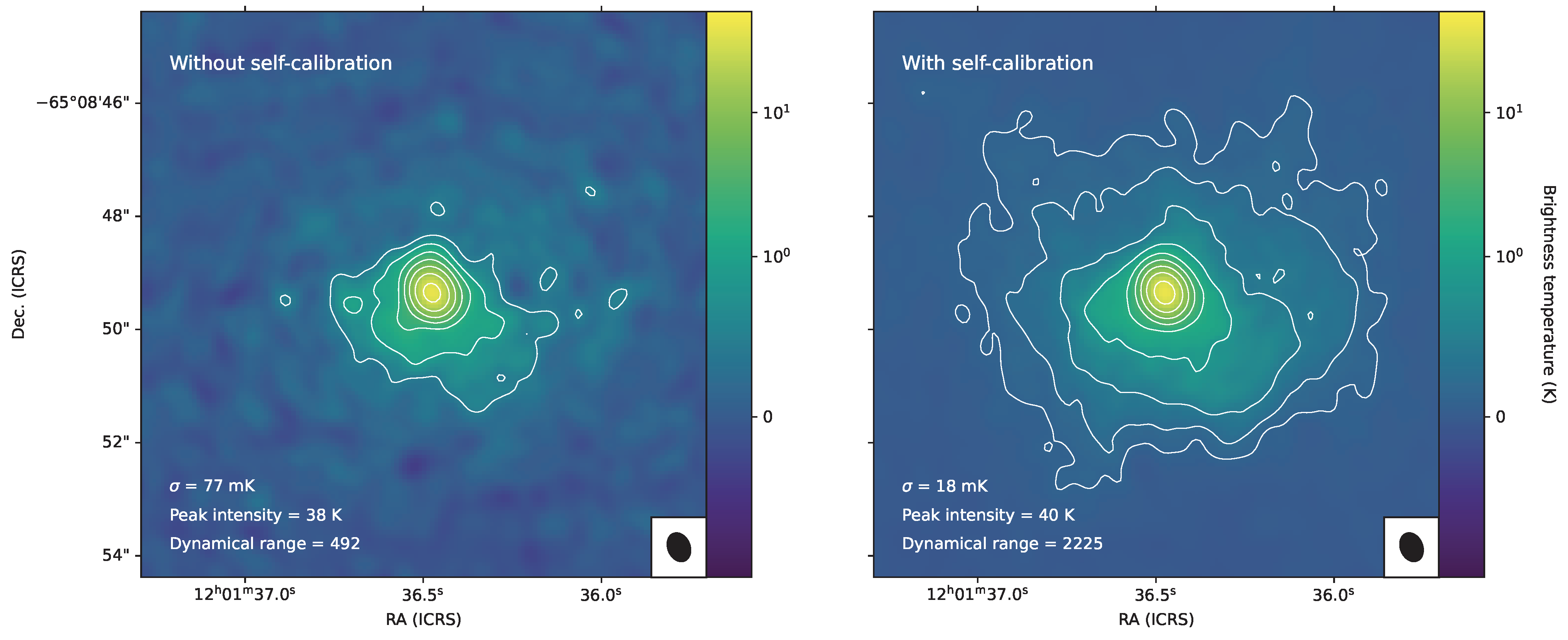}
        \caption{Comparison between the continuum emission at 279.5~GHz in BHR71-IRS1 with (right panel) and without applying self-calibration (left panel). The color maps show the brightness temperature. The color scale is linear between -1~K and 1~K and logarithmic for larger values, in order to enhance the contrast for low intensities. In each panel, the contours start at 6$\sigma$, and then increase by a factor 2 (i.e., 12$\sigma$, 24$\sigma$, 48$\sigma$ and so on). The values of $\sigma$, peak intensity and dynamical range (i.e., the peak intensity divided by $\sigma$) are given in the bottom left corner of each panel. The black ellipse in each panel represents the synthesized beam.\label{fig:bhr71-selfcal-comparison}}
    \end{minipage}
    \hfill 
    \begin{minipage}[t]{0.34\textwidth}
        \centering
        \includegraphics[width=\textwidth]{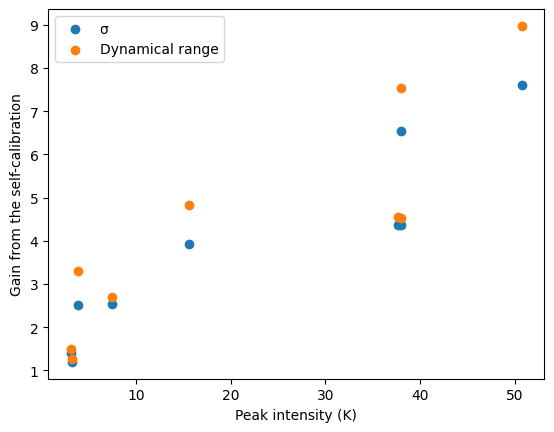}
        \caption{Gain of the self-calibration on the continuum images at 279.5~GHz in all sources except NGC 1333-SVS13A and L1551-IRS5, as a function of the peak intensity. Blue points show the ratio of the noise level in the continuum image without and with self-calibration applied. Orange points show the ratio between the dynamical range of the continuum image with and without self-calibration. \label{fig:selfcal-gain}}
    \end{minipage}
\end{figure*}

In this section, we discuss in more details the improvement of the
continuum images by the phase self-calibration technique. In
Fig.~\ref{fig:bhr71-selfcal-comparison}, we compare the continuum
emission in BHR71-IRS1 at 279.5~GHz (which corresponds to SPW 25 of tuning 1), with and without applying
self-calibration. Both maps were deconvolved using the same stopping
criteria, based on convergence of the total flux in the deconvolved
image. The image without self-calibration has apparent side lobes with
negative emission. These side lobes are significantly reduced by the
self-calibration. In addition, the self-calibration allows recovering
the relatively extended low-intensity emission around the continuum
peak. For BHR71-IRS1, as an example, the self-calibration decreases the noise by a
factor of 4.3, and it increases the dynamical range by approximately the
same factor (4.5). Self-calibration therefore leads to a significant
improvement in terms of image quality, noise level and dynamical
range.

To check whether similar improvements were obtained for other sources, we measured the gain of the self-calibration, in terms of noise and dynamical range, on the 279.5~GHz emission in all the sources of the sample. In Fig.~\ref{fig:selfcal-gain}, we show the gains as a function of the peak continuum intensity, for each source. SVS~13 and L1551-IRS5 are not included in this figure because these two sources were observed with a different array configuration. The gains on the noise are similar to the gains in dynamical range for any given source. We find that, at first order, the gains depend on the source peak intensity. For the weakest source, the self-calibration reduces the noise by a factor 1.2, and it improves the dynamical range by a factor 1.3. For the brightest source, it reduces the noise level by a factor 7.6, and it increases the dynamical range by a factor 9.0. This finding is consistent with the continuum observation being limited by the dynamical range. The brightest sources are obviously the most affected by this limitation, and the improvement that the self-calibration brings is larger for the brightest sources. Overall, we conclude that the self-calibration improves the continuum images
for all sources, with the largest noise reduction and dynamical range increase (by almost an order of magnitude) obtained for the brightest sources.

\end{appendix}
\end{document}